\documentclass[10pt,twocolumn,letterpaper]{article}

\usepackage{iccv}              

\definecolor{iccvblue}{rgb}{0.21,0.49,0.74}
\usepackage[pagebackref,breaklinks,colorlinks,allcolors=iccvblue]{hyperref}
\usepackage{amsmath,amssymb,amsfonts,mathrsfs}
\usepackage{multirow,tabularray}
\usepackage{CJKutf8} 
\usepackage{MnSymbol,wasysym,fontawesome5}
\usepackage[accsupp]{axessibility}  

\def\paperID{10398} 
\def\confName{ICCV}
\def\confYear{2025}

\newcommand{\mbf}[1]{\mathbf{#1}}

\title{Coordinate-based Speed of Sound Recovery for Aberration-Corrected Photoacoustic Computed Tomography}

\author{Tianao Li$^{1,4}$ \quad Manxiu Cui$^2$ \quad Cheng Ma$^3$ \quad Emma Alexander$^{1,4}$ \\
$^1$Northwestern University \quad
$^2$California Institute of Technology \quad
$^3$Tsinghua University \\
$^4$NSF-Simons AI Institute for the Sky (SkAI) \\
{\tt\small tianaoli@u.northwestern.edu}
}

\begin{document}
\twocolumn[{
\maketitle
\begin{center}
    \includegraphics[width=\linewidth]{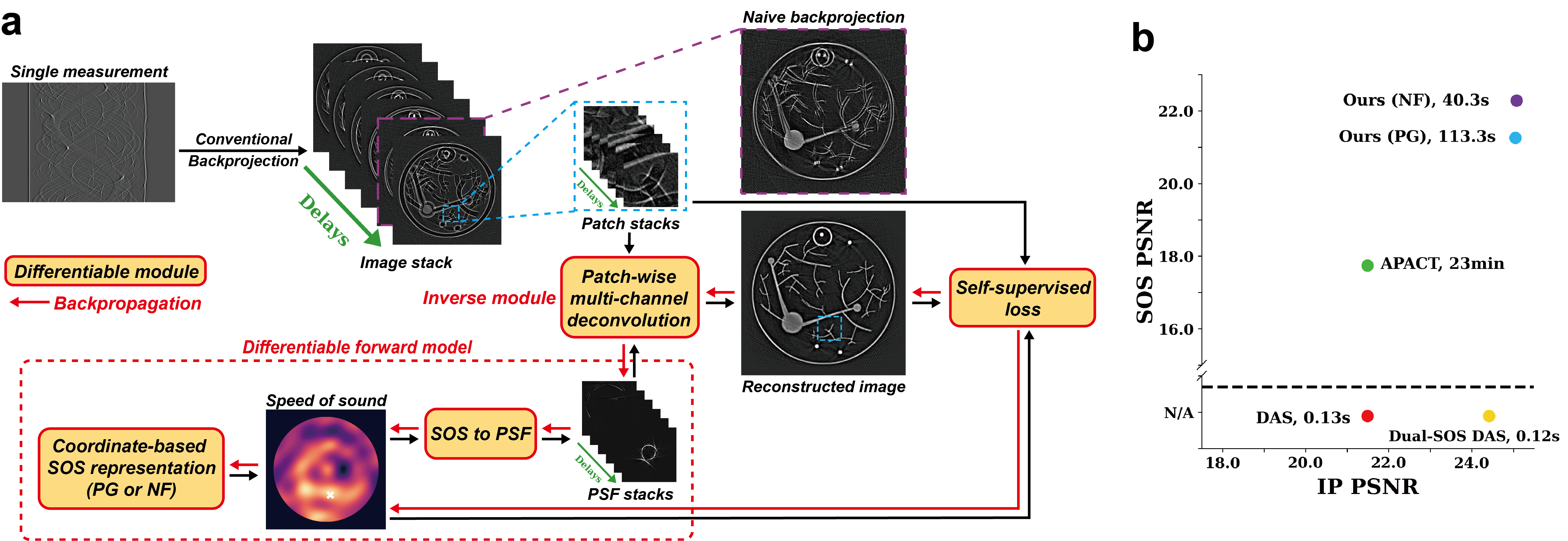}
    \captionof{figure}
    {
    \textbf{Self-Supervised Speed of Sound Recovery Corrects PACT Aberrations.} (a) Backprojection converts photoacoustic measurements into images (purple), which suffer aberrations caused by unaccounted-for variation in the speed of sound (SOS) of tissue. Our method produces high-quality images through a self-supervised recovery of SOS. We begin with a stack of aberrated reconstructions, created with a varying delay parameter (green), roughly analogous to a focal stack. Each image patch is recovered with a multi-channel deconvolution. Our forward model (dashed red) uses a trainable coordinate-based SOS representation to estimate the point spread functions (PSFs) at each location (white ``x" for example shown). The physics-based forward model and multi-channel inversion are fully differentiable, enabling test-time training without any external training data, which is currently limited for this imaging modality. (b) We benchmark our method using both a neural field (NF) and pixel grid (PG) for SOS representation. Compared to the best existing methods, both with and without SOS recovery, we provide state-of-the-art image quality, and recover SOS with an order-of-magnitude speed-up.
    }
    
    \label{fig:overview}
\end{center}
}]

\maketitle
\begin{abstract}

Photoacoustic computed tomography (PACT) is a non-invasive imaging modality, similar to ultrasound, with wide-ranging medical applications. 
Conventional PACT images are degraded by wavefront distortion caused by the heterogeneous speed of sound (SOS) in tissue. 
Accounting for these effects can improve image quality and provide medically useful information, but measuring the SOS directly is burdensome and the existing joint reconstruction method is computationally expensive. 
Traditional supervised learning techniques are currently inaccessible in this data-starved domain. 
In this work, we introduce an efficient, self-supervised joint reconstruction method that recovers SOS and high-quality images for ring array PACT systems. 
To solve this semi-blind inverse problem, we parametrize the SOS using either a pixel grid or a neural field (NF) and update it directly by backpropagating the gradients through a differentiable imaging forward model. 
Our method removes SOS aberrations more accurately and 35x faster than the current SOTA. 
We demonstrate the success of our method quantitatively in simulation and qualitatively on experimentally-collected and in vivo data. 
Our code and synthetic numerical phantoms are available on our project page: \href{https://lukeli0425.github.io/Coord-SoS-PACT/}{\tt \small https://lukeli0425.github.io/Coord-SoS-PACT/}.
\end{abstract}    
\section{Introduction}
\label{sec:intro}
Photoacoustic computed tomography (PACT) images biological samples without cutting through tissue or disrupting normal function~\cite{xia2014photoacoustic}. Biological tissues absorb energy from laser stimulation and emit ultrasonic waves, which propagate through the rest of the sample and are then collected by an external array of ultrasound transducers. An image of the initial pressure (IP) distribution (2D for the scope of this paper) is obtained by processing the ultrasonic signals collected by the transducer array. Traditional image reconstruction algorithms include filtered back-projection~\cite{xu2005universal} and delay-and-sum (DAS)~\cite{thomenius1996evolution, hoelen2000image, jeon2019real}, which back-project the collected signals into the image using times of flight assuming a constant SOS throughout the sample.
This assumption neglects the impact of the sample's internal structure and leads to image aberrations. Specifically, the inherent acoustic heterogeneity, or spatial variations in the 2D SOS distribution, will cause wavefront distortions in the ultrasonic waves that speed or delay their arrival to different parts of the transducer array. 
Visual artifacts of these distortions include ringing, smear, and doubling of features, which can obscure key sample structures~\cite{xu2003effects}. 


Though supervised learning methods have made significant progress on image restoration in many applications, PACT is a relatively novel imaging modality and appropriate datasets (large and naturally distributed set of measurements, with realistic aberrations, accompanied by ground truth for both image and SOS values) are not available. We consider a {\em ring array} PACT system and address these challenges with a \textit{self-supervised} joint recovery method that learns SOS by backpropagation through a differentiable physical model, then uses the SOS to reconstruct a clean image.
Specifically, as shown in in~\cref{fig:overview}, we optimize a differentiable coordinate-based representation of the SOS, parameterized by a pixel grid (PG) or neural field (NF), to best match observed aberrations, which can then be removed with deconvolution. By imposing physical consistency across image locations and a delay-based ``focal stack" (details in \cref{sec:method}), this method converges quickly to produce an approximate SOS and high quality image.
Our key contributions are:

 
1. A physics-based and interpretable PACT imaging model that is fully differentiable, enabling time- and memory-efficient optimization through backpropagation. 

2. A more accurate, globally consistent representation of wavefront errors, for SOTA reconstruction quality in both image and SOS.

3. A self-supervised, test-time training framework that requires no external training data and can flexibly incorporate novel SOS representations and regularizers. 

We benchmark the proposed method and previous methods on simulated data, showing that our method achieves state-of-the-art reconstructions in a fraction of the compute time (40 seconds vs. 23 minutes).  In qualitative results on real data, we show the method is robust to nonideal effects (e.g. transducer responses and geometric offsets) found in PACT systems, as demonstrated by its translation to {\em in vivo} data without modification.

\section{Related Work}
\label{sec:related_work}

\paragraph{Direct SOS Aberration Mitigation} 
A wide range of methods have been developed to mitigate or remove the artifacts caused by acoustic heterogeneity in PACT, as recently reviewed in~\cite{wang2020combating, tang2023advanced}. Autofocus approaches optimize an assumed-constant SOS based on the reconstructed image quality~\cite{treeby2011automatic} or the coherence factor of the photoacoustic (PA) signals~\cite{yoon2012enhancement,cong2015photoacoustic}. Half-time reconstruction methods~\cite{anastasio2005half, poudel2017mitigation} likewise assume a constant SOS but alleviate aberrations by identifying and discarding regions of the signals that are more affected by acoustic heterogeneity. Among these methods, Dual-SOS DAS~\cite{li2017single} provides the best image reconstruction quality, by efficiently correcting the first-order effect of acoustic heterogeneity by representing the sample with an oval-shaped region of altered SOS and assuming values for its shape and internally-constant SOS. On the other end of the complexity scale, Ultrasound Computed Tomography (USCT)-enhanced methods~\cite{jin2006thermoacoustic,manohar2007concomitant,xia2013enhancement} directly measure the 2D SOS with a complementary ultrasound measurement apparatus. This measurement enhances image reconstruction but is challenging and expensive to implement simultaneously with PACT.
The above methods either require oversimple assumptions or costly measurement of the SOS, and can only unreliably mitigate instead of fully undoing the SOS aberration in practice~\cite{wang2020combating}.

\paragraph{Joint Recovery of SOS and Image}
In contrast to methods that assume or measure SOS, joint reconstruction approaches simultaneously recover images and 2D SOS from PA signals. Regularization-based joint reconstruction~\cite{zhang2006reconstruction, shan2019simultaneous} alternatively reconstructs the image and SOS in an iterative framework with regularization on both distributions. However, the ill-posed nature of the problem causes numerical instability and error accumulation. The feature coupling method~\cite{cai2019feature} parameterizes the SOS with several constant value regions and maximizes the similarity between reconstructed images from partial arrays, but is unable to model a complicated SOS map and relies heavily on a good initial guess. Inspired by indirect wavefront sensing in adaptive optics (AO)~\cite{ji2017adaptive,hampson2021adaptive,zhang2023adaptive}, Adaptive PACT (APACT)~\cite{cui2021adaptive} divides the image into small patches and solves for the wavefront in every patch via an exhaustive search. Of these methods, APACT produces state-of-the-art image quality, but with two major drawbacks. 
First, APACT assumes a simplified wavefront model that cannot account for overall scaling of the reconstructed image. 
Second, its exhaustive search is computationally expensive, leading to extended reconstruction times that limit its practical application. 
Our method addresses these weaknesses with a novel physics-based, self-supervised learning approach.

\paragraph{Supervised Learning}
Deep learning has been applied to SOS aberration correction~\cite{jeon2020deep,jeon2021deep} and other image restoration tasks~\cite{deng2021deep} in PACT. However, a key challenge lies in generalizing models trained on large simulated datasets to real-world \textit{in vivo} data~\cite{tang2023advanced}. The scarcity of realistic, high-fidelity simulated data causes a domain shift between training and testing that can significantly degrade results on real data. Our physics-based framework, in contrast, shows robust performance across both simulated and \textit{in vivo} data. 


\paragraph{Neural Fields}
Neural fields (NFs), or implicit neural representations (INRs) have been
shown to be capable of representing a variety of natural signals~\cite{sitzmann2020implicit, xie2022neural}, and have been demonstrated in many computational imaging tasks, including medical imaging~\cite{molaei2023implicit, reed2021dynamic, zang2021intratomo, sun2021coil, shen2022nerp, xu2023nesvor}. 
NFs, as well as interpolated pixel grids~\cite{fridovich2022plenoxels}, represent images with no need for external or labeled training data (see comparisons in~\cite{fridovich2022plenoxels, kim2025grids}). 
%
Recent works in PACT have utilized NFs for IP image reconstruction from limited or sparse measurements~\cite{zou2024pa, xiao2024unsupervised, yao2024implicit}, but these assume constant SOS and only reconstruct the IP image. NFs have also been used to reconstruct SOS in ultrasound CT~\cite{byra2024implicit} and have shown promise in estimating wavefronts for computational image reconstruction~\cite{kang2024coordinate} and closed-loop control systems for adaptive optics~\cite{feng2023neuws}. 
We, like \cite{kang2024coordinate}, do not intervene in the measurement process, but computationally correct for wavefront-based aberration parameters estimated with a NF. We use the SIREN NF architecture, which is a fully connected network with sinusoidal activations~\cite{sitzmann2020implicit}.

\section{Method}
\label{sec:method}

\subsection{Image Formation and Aberration}
Filtered back-projection~\cite{xu2005universal} is a widely used image reconstruction method in ring array PACT systems. In practice, due to the limited bandwidth and non-flat angular response of the transducers~\cite{cui2021adaptive}, a delay-and-sum (DAS) algorithm~\cite{thomenius1996evolution, hoelen2000image, jeon2019real} is often used as a reasonable approximation. A delay-and-sum image $y(\mbf{r}; d)$ varies over sample locations $\mbf{r}'$ as well as a user-selected extra delay parameter $d$ that functions as a ``focus setting" \cite{treeby2011automatic}, and is reconstructed according to
\begin{equation}
    y(\mbf{r}'; d) = \sum_{n=1}^{N_{t}} S \left(t=\frac{\vert \mbf{r}' - \mbf{r}_n \vert - d}{v_0}, n\right),
    \label{eq:DAS}
\end{equation}
where $v_0$ is the assumed constant SOS, $S(t,n)$ is the PA signal at time $t$ for the n$^{\text{th}}$ transducer, $\mbf{r}_n$ is the location of the n$^{\text{th}}$ transducer, and $N_t$ is the total number of transducers.

\begin{figure}[!t]
    \centerline{\includegraphics[width=\columnwidth]{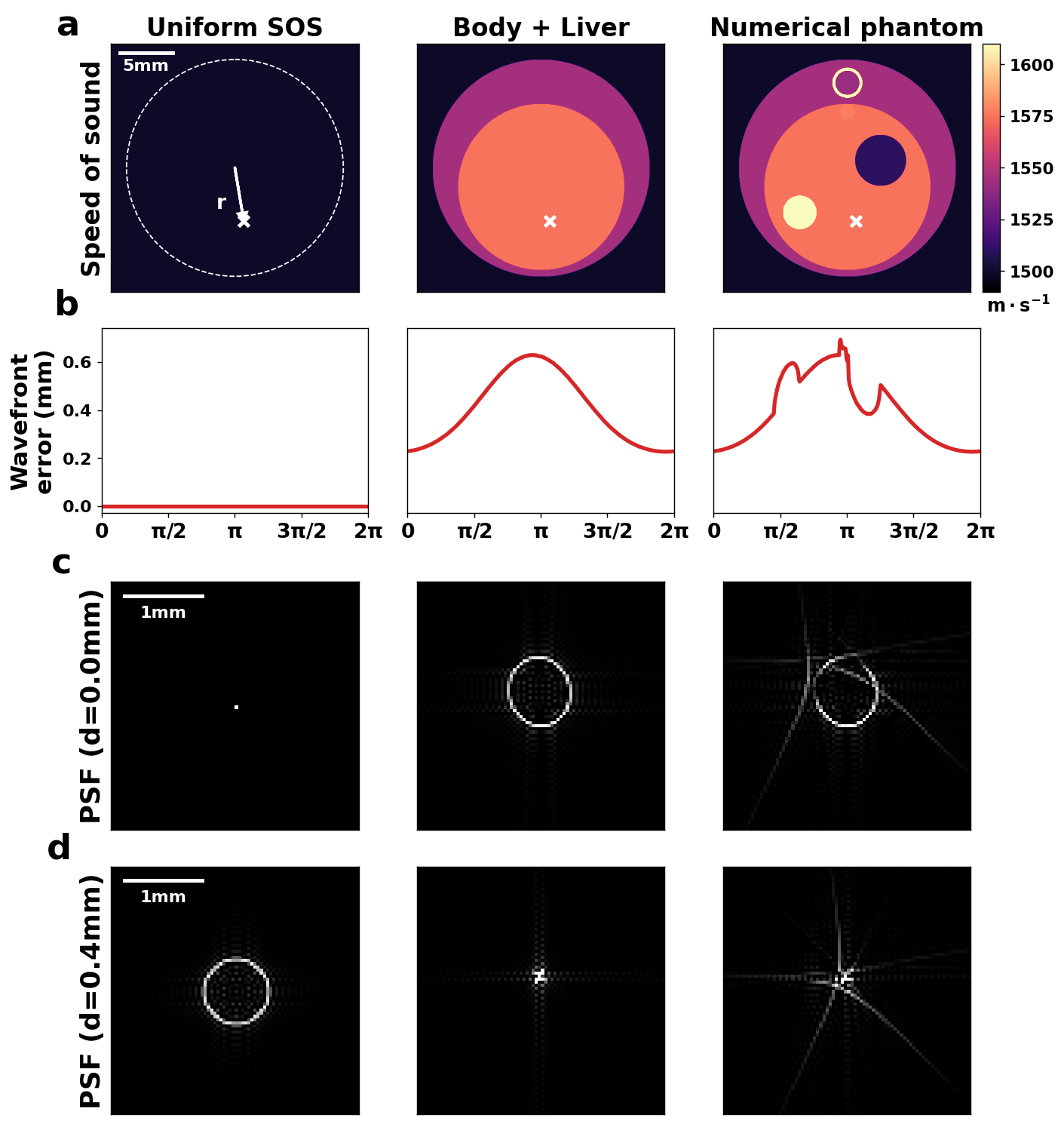}}
    \caption{\textbf{Acoustic heterogeneity creates non-ideal PSFs due to wavefront distortions.}
    We consider three SOS patterns (a): the uniform SOS assumed by standard methods, a simplified body model with two internal SOS values, and a more realistic numerical phantom that we created to mimic a body slice including representative materials ranging from air to bone. For the location marked with a white ``x", we illustrate (b) the discrepancy between the standard methods' assumed wavefront and the true wavefront location arising from the SOS shown. These wavefront errors determine the PSFs (c) at the indicated location for each SOS, which lead to image aberrations for non-pointlike PSFs. A user-set delay value in the image reconstruction changes the resulting PSFs and aberrations (d), and can be used to effectively remove aberrations for simple SOS models but not realistic ones.
    }
    \label{fig:SOS}
\end{figure}

If the SOS throughout the sample is constant and matches the background, there is a single velocity $v_0$ and delay $d=0$ that will create a sharp image. Because this is not the case for real samples, points in the object are smeared across the image into irregular patterns~\cite{xu2003effects}.
The delay value $d$ is a mitigation for these aberrations, and can improve image quality bending each back-projected arc, with the risk that poorly-selected delays will spread sample points into rings on the image (see PSFs in first column of Fig.~\ref{fig:SOS}). In real samples, with heterogeneous SOS, more complex aberrations arise and cannot be resolved with a single delay parameter (compare to right columns of Fig.~\ref{fig:SOS}).
Specifically, assuming a straight acoustic ray model~\cite{xu2003effects}, the time-of-flight of a signal from sample point $\mbf{r}'$ to a transducer at $\mbf{r}$ is
\begin{equation}
    t(\mbf{r}', \mbf{r}, \mbf{v}) = \int_{\mbf{r}'}^{\mbf{r}} \frac{1}{v(\mbf{l})} dl,
    \label{eq:TOF}
\end{equation}
where $\mbf{v}$ is the unknown 2D SOS. 
From~\cite{cui2021adaptive}, the angle-varying displacement between the actual, distorted wavefront at location $\mbf{r}'$ for SOS $\mbf{v}$ and the uniform wavefront assumed by DAS is the wavefront error $w(\theta;\mbf{r}',\mbf{v})$:
\begin{equation}
    \begin{aligned}
        w(\theta; \mbf{r}', \mbf{v}) &= \Vert \mbf{r}' - \mbf{r}(\theta) \Vert - t(\mbf{r}', \mbf{r}(\theta), \mbf{v}) v_0 
        \\
        &= \int_{\mbf{r}'}^{\mbf{r}(\theta)} \left(1 - \frac{v_0}{v(\mbf{l})} \right) dl,
    \end{aligned}
    \label{eq:wf}
\end{equation}
where $\theta$ is the angle of the vector $\mbf{r}-\mbf{r}'$. See Fig.~\ref{fig:SOS}b for example wavefront errors corresponding to the SOS maps $\mbf{v}$ shown in Fig.~\ref{fig:SOS}a for the location $\mbf{r}'$ indicated by a white ``x". 
It has been shown in~\cite{cui2021adaptive} that the resulting aberrations, despite varying across the sample, can be modeled within isoplanatic image patches $y_i$ as convolutions with PSFs $h_i$ and unaberrated patches $x_i$:
\begin{equation}
    y_i(\mbf{r}';d,\mbf{v}) = h_i(\mbf{r}';d,\mbf{v}) \ast x_i(\mbf{r}').
    \label{eq:conv}
\end{equation}
Specifically, these PSFs $h_i$ can be computed in the Fourier domain (indicated with capital letters) from the wavefront-based transfer function $H=\mathscr{F}(h)$ as
\begin{equation}
    \begin{aligned}
        H_i(\mbf{k};d,\mbf{v}) 
        &= \frac{1}{2} \left( e^{-j|
        \mbf{k}|(d - w(\angle\mbf{k};\mbf{r}'_i,\mbf{v}))} + e^{j|\mbf{k}|(d - w(\angle\mbf{k}+\pi;\mbf{r}'_i,\mbf{v}))}\right),
        \label{eq:TF}
    \end{aligned}
\end{equation}
where $\mbf{k}$ is the frequency coordinate and $\mbf{r}'_i$ is the center location of the i$^\text{th}$ image patch. See Fig.~\ref{fig:deconv}a for an illustration of PSFs at varying locations $\mbf{r}'$ and delays $d$ in our numerical phantom. At each location, the delay that generates the most compact PSF is indicated in red.

\subsection{Multi-channel Deconvolution}

\begin{figure}[!t]
    \centerline{\includegraphics[width=\columnwidth]
    {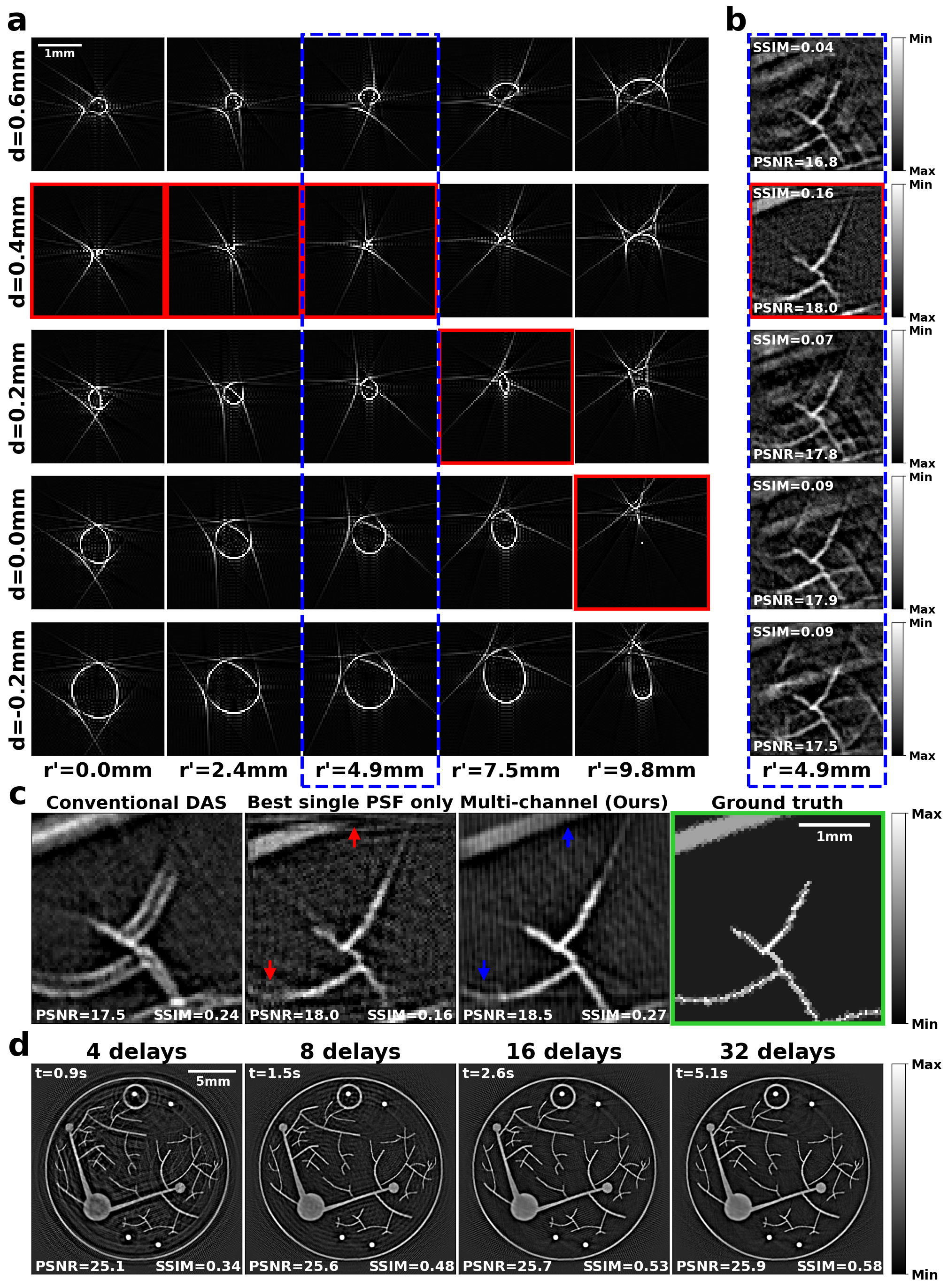}}
    \caption{\textbf{Multi-channel Deconvolution.} 
    Our method derives robustness by predicting aberrations across multiple delay values and at all image patches. In (a) we visualize PSFs computed from one of our numerical phantoms and image aberration model. For each image patch (rows, parameterized by radial distance $r'$ from image center along white arrow shown in Fig.~\ref{fig:SOS}a) we compute a stack of PSFs that correspond to a range of delay distances (columns, 5 of 32 shown). 
    In each column, the most concentrated PSF (hence the easiest to deconvolve) is marked with a red border. We pick one image location (blue outline) and demonstrate in (b) that the single-channel deconvolved image patch reaches the highest quality when using the most compact PSF (red) but still suffer from significant artifacts (compare to ground truth outlined in green beneath). We show in (c) that these artifacts can be mitigated with multi-channel deconvolution, including the recovery of lost features at patch edges (arrows) due to shifts in PSF centers across delays. We show in (d) the improvement in image quality as well as the growth of compute time for our multi-channel deconvolution with an SOS oracle as delay channels are added. 
    All other results shown use 16 delays by default. 
    }
    \label{fig:deconv}
\end{figure}

Joint processing across multiple delay values $\{d_j\}_{j=1,\dots,M}$ provides more robust deconvolution results. 
A DAS reconstruction for each extra delay distance comprises our DAS stack, 
and for each patch of this image stack these delay values (in conjunction with SOS $\mbf{v}$, on which more later) generate a corresponding PSF stack. Because each PSF takes the same underlying clean image into the aberrated image stack, the problem is {\em overdetermined} and can be expressed in the frequency domain as
\begin{equation}
    \left[ \begin{array}{cc}
       Y_i(\mbf{k}; d_1, \mbf{v}) \\
       \vdots \\
       Y_i(\mbf{k}; d_{M},\mbf{v}) \\
    \end{array} \right]
     = 
     \left[ \begin{array}{cc}
       H_i(\mbf{k}; d_1,\mbf{v}) \\
       \vdots \\
       H_i(\mbf{k}; d_{M}, \mbf{v}) \\
    \end{array} \right]
     X_i(\mbf{k}),
    \label{eq:fm_fourier}
\end{equation}
and in matrix form as $\mbf{Y}_i = \mbf{H}_i(\mbf{v}) X_i$.
We reconstruct clean image patches $\hat{X}_i(\mbf{Y}_i,\mbf{H}_i(\mbf{v}))$ with a pseudo-inverse,
\begin{equation}
    \centering
    \begin{aligned}
        \hat{X}_i(\mbf{Y}_i,\mbf{H}_i(\mbf{v})) 
        & = \frac{\mbf{H}_i(\mbf{v})^{\top} \mbf{Y}_i}{\mbf{H}_i(\mbf{v})^{\top}\mbf{H}_i(\mbf{v})}.
    \end{aligned}
    \label{eq:multi-deconv}
\end{equation}
This simple method succeeds thanks to several sources of robustness: the combination of information encoded across multiple delay channels,  
the fact that the digital creation of new channels generates no additional measurement error, and later in our pipeline the consistency across overlapping image patches through Gaussian-weighted merging. 

Fig.~\ref{fig:deconv}b shows the image patch recovered using single-channel deconvolution with each delay's PSF. The most compact PSF (red) produces the best result of these, but artifacts are extreme for each individual channel. By using 16 delays for multi-channel deconvolution, our method reduces ringing and includes features that are otherwise lost at patch edges, see arrows in Fig.~\ref{fig:deconv}c. The reduction of aberrations as more delay channels are added can be seen in Fig.~\ref{fig:deconv}d, which also shows a growth in compute time from 0.9 to 5.0 seconds for 4 to 32 channels. 

\begin{figure*}[!t]
    \centerline{\includegraphics[width=2\columnwidth]{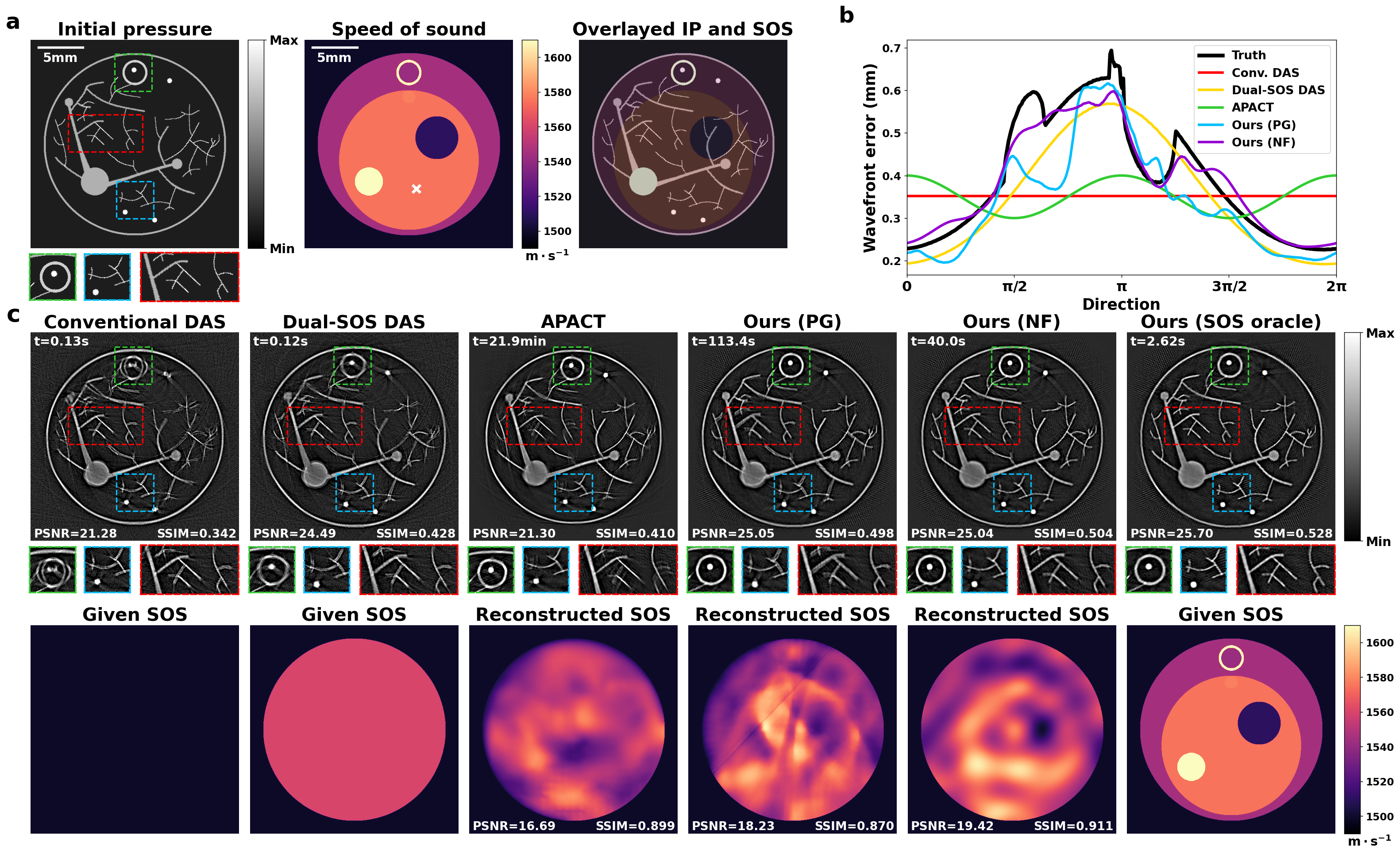}}
    \caption{\textbf{Numerical simulation demonstrates accuracy and computational efficiency.} (a) Our numerical phantom allows quantitative evaluation of image reconstruction under different SOS distributions in simulation. The PA signals are generated with k-Wave simulation \cite{treeby2010k} using the IP and SOS shown. 
    (b) The wavefront (sampled location marked by white ``x" in a) reconstructed by our method with NF (purple) and PG (green) is much closer to the true wavefront (red) than
    other methods, better capturing image aberrations. 
    (c) We compare our method with conventional DAS, Dual-SOS DAS, our multi-channel deconvolution with the true SOS, and APACT. While conventional DAS assumes a uniform SOS and results in significant image aberrations, Dual-SOS DAS provides better contrast by assuming a single-body SOS but cannot fully undo the aberrations and reconstruct the SOS. Given an accurate SOS, our multi-channel deconvolution is able to efficiently reconstruct a high-quality IP image. APACT is able to deblur the image effectively but does not address overall shrinkage (see vertical shift in green boxes) and cannot reconstruct an accurate SOS. Our method with NF solves the shrinking problem by fully characterizing the wavefront and is able to fully reconstruct the IP and SOS (see metrics) in a much shorter time than APACT. 
    Ours with PG also produces good IP images but generates stripe-like artifacts in the SOS that originate from the path-integral nature of the wavefront computation. Providing our method with the ground truth SOS leads to only marginal improvements in image quality.
    See visualizations of the convergence in the supplementary videos.
    }
    \label{fig:numerical}
\end{figure*}

\subsection{Joint Reconstruction}
\label{sec:jr}
By combining our differentiable PSF generation and deconvolution modules, shown in red in Fig.~\ref{fig:overview}, we are able to backpropagate through the entire forward and inverse models to achieve consistency between the expected and observed image aberrations.

Note in~\cref{eq:wf} that our wavefronts are 
calculated from the entire SOS map via path integral, therefore each incorrect SOS pixel will contribute to an incorrect PSF at \textit{every} image patch through the wavefront in some direction. 
While the PSFs will be included in both the forward and inverse model, the overdetermined nature of~\cref{eq:fm_fourier} brings the PSFs into consistency with each other as iterations progress, enabling joint reconstruction of the unaberrated image and the SOS from PA signals alone.


We benchmark two types of SOS representation: an interpolated pixel grid (PG) with the same resolution as the IP image, and a neural field (NF).
For the latter, we used a single-layer, 256-feature, fully-connected network with sinusoidal activations~\cite{sitzmann2020implicit}. This non-pretrained, coordinate-based network maps pixel coordinates to SOS predictions. 
The total number of NF parameters (1027) is much smaller than the number of pixels ($\sim$200k) in the SOS map, which  reduces the degrees of freedom in the inverse problem and acts as an implicit regularizer that makes the problem better-posed. 
See~\cref{sec:ablation_network} for ablations on network size.
For both representations, we reduce unnecessary computation by restricting these pixel grid or NF coordinates to a circular mask roughly the size of the sample and fix the background SOS based on measured water temperature~\cite{marczak1997water}.

We train the SOS parameters $\phi$ to minimize a self-supervised aberrated-image-matching loss:
\begin{equation}
    L(\mbf{v}_\phi) = \sum_{i=1}^N |\mbf{k}|\left\Vert \mbf{Y}_i - \mbf{H}_i(\mbf{v}_\phi)\hat{X}_i(\mbf{Y}_i, \mbf{H}_i(\mbf{v}_\phi))\right\Vert_2^2 +\lambda \text{TV}(\mbf{v}_\phi),
     \label{eq:jr}
\end{equation}
over all $N$ image patches simultaneously, where $\phi$ is the parameters in the PG or NF. This loss function imposes a physics-informed data fidelity that includes both the forward model in $\mbf{H}_i$ as well as a Fourier-noise-specific $|\mbf{k}|$ weighting as in~\cite{cui2021adaptive}. 
We use TV loss to regularize the conventional pixel grid representation, while for the NF representation, we rely instead on the network's implicit regularization and set $\lambda=0$.
Training is performed only on the SOS representation parameters $\phi$ but requiring accurate aberration reconstruction across all delays and image patches leads to sufficiently accurate PSFs that the final multi-channel deconvolution produces high-quality recovered IP image patches $x_i$. Finally, these patches are merged into a single image by applying a Gaussian window with a FWHM of 1.5 mm~\cite{cui2021adaptive} for smooth interpatch transitions and adding the patches back to the IP image at corresponding patch locations.
We use a delay count of $M=16$ by default in this paper (see ablations in~\cref{sec:ablations}).
See~\cref{sec:image_stichting} in the supplement for details on our image patches and stitching.

Our method is implemented with PyTorch~\cite{paszke2019pytorch} and available on GitHub (see project page). We trained for 10 epochs for NF and 30 epochs for PG with Adam optimizer~\cite{kingma2014adam} and learning rates of $5\times10^{-3}$ and $1\times10^{-1}$ respectively on an NVIDIA RTX A6000 GPU. 


\section{Results}
\label{sec:results}
\subsection{Numerical Simulations}

To quantitatively benchmark the performance of our method with ground truth references, we designed a set of numerical phantoms with 5 different SOS maps (see~\cref{sec:simulations_supp} for details), one of which is shown in Fig.~\ref{fig:numerical}a for visualization. 

The wavefront errors modeled/reconstructed (sample location marked by white ``x" on SOS in Fig.~\ref{fig:numerical}a) by different methods are shown in Fig.~\ref{fig:numerical}b. To use the terminology of~\cite{cui2021adaptive}, the wavefront errors can be decomposed into a Fourier series, where the components degrade the IP image differently.
Conventional DAS assumes a uniform SOS thus a constant wavefront (red), while Dual-SOS DAS assumes an additional body SOS and is able to capture the zeroth and first-order term accurately (yellow). 
APACT (green) includes only the zeroth and second order Fourier series terms in its wavefront model (see~\cite{cui2021adaptive} and~\cref{sec:apact_wavefront} in the supplement for details), missing both the first order (the cosine component captured by dual-SOS in yellow) as well as higher order terms capturing finer detail. In comparison, our methods can accurately model lower and higher-order components in the wavefront (purple and blue), compared with the truth wavefront (black). 




These wavefront errors are directly reflected in image reconstruction quality in~\cref{fig:numerical}c. Conventional DAS, which assumes a uniform SOS of 1510.0 m/s, not only suffers from doubling artifacts (see insets) due to the ignorance of higher order terms but also causes the image to shrink (see vertical shift of the circle and boundary in green box). This shrinking is caused by the missing first-order term in the wavefront, which leads to off-center PSFs that shift the image patches toward the center of the tissue. Note also that for real samples this uniform SOS value is tuned by generating many reconstructions and requiring the user to select the best image, significantly increasing the reconstruction time in practice.
Dual-SOS DAS assumes a body SOS of 1561.0 m/s (also hand-tuned with search time not included in reported time) and is able to fix the shrinking problem but artifacts remain. 
The image reconstructed by APACT is also shrunk (see green box and the image quality metrics) due to ignorance of the first-order term. APACT oversmooths the SOS and loses key features in the SOS, e.g., the orange liver-like region and the hole in the upper right, and is also slow and expensive due to the exhaustive search for wavefronts coefficients. 
Our method reconstructs a more accurate SOS, corresponding to a significant improvement in image quality over traditional methods. Image patches show fewer aberrations than all previous methods, reflected in high PSNR and SSIM scores. Our method with NF takes only 40 seconds, compared to 23 minutes for APACT. Our method with PG produces decent IP images in similar compute time but suffers from stripe-like SOS artifacts that originate from the path-integral nature of the wavefront calculation (see~\cref{eq:wf}) in the straight acoustic ray model.
We also show the image reconstructed by multi-channel deconvolution using the ground truth SOS (SOS oracle) as an upper limit baseline for image quality and speed of USCT-enhanced reconstruction. 
See~\cref{tab:results} for quantitative evaluation of all the methods, averaged over five numerical phantoms.
Supplementary videos show the convergence of the IP image and SOS for each experiment.


\begin{figure*}[!t]
    \centering
    \includegraphics[width=1.9\columnwidth]{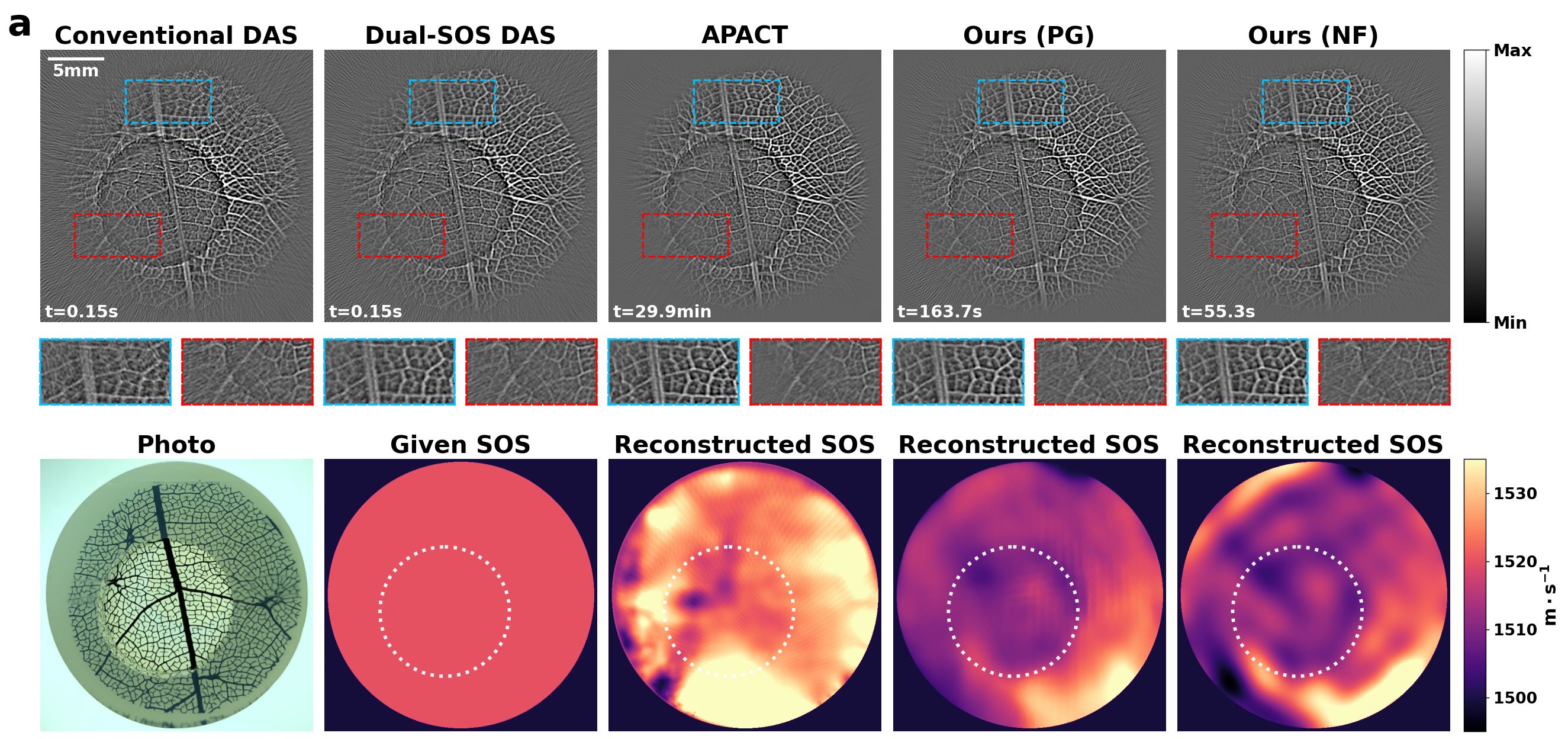}
    \includegraphics[width=1.9\columnwidth]{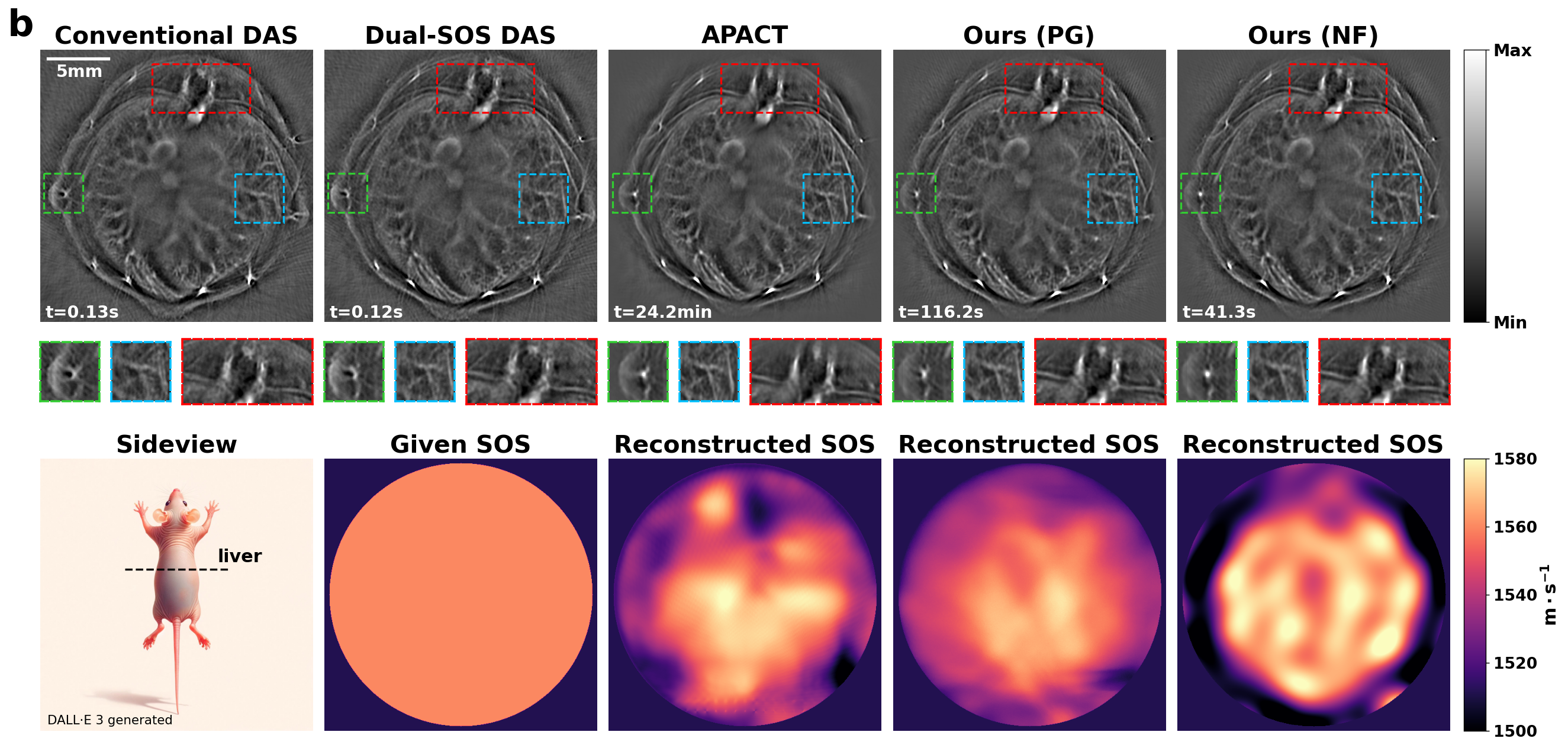}
    \caption{\textbf{Real-world results.} We compare our methods to conventional DAS, Dual SOS DAS, and APACT on experimentally collected data, for which the true IP and SOS are unknown. 
    Panel (a) shows a 3-SOS leaf-and-gel phantom~\cite{cui2021adaptive} (see labeled photograph). 
    Prior methods fail to capture the vein structure accurately, while our method succeeds in both simple regions (blue) as well as more challenging regions (red) featuring dim illumination and a material boundary. Our SOS map approximately recovers the material change in the phantom (white dotted line).  
    Panel (b) shows the reconstructed images from \emph{in vivo} mouse liver data from~\cite{cai2019feature}. Despite requiring less computational time, our method precisely locates features like bright blood vessels and body edges (green) and effectively recovers fine structures with high contrast (blue and red). Notably, the SOS map reconstructed by our method with NF exhibits a superior match to the liver's anatomical shape compared to APACT (see~\cref{fig:overlay} in supplement for overlaid SOS and IP). In both experiments, our method with PG produces artifacts in the SOS despite constrained by a strong TV regularization. See supplementary videos for visualizations of convergence.} 
    \label{fig:real}
\end{figure*}

\begin{table}[ht]
    \small
        \setlength{\tabcolsep}{3.5pt}
        \centering
        \begin{tabular}{@{}l|c|c|c|c|c@{}}
            \toprule
            \bf Method & \begin{tabular}{@{}c@{}} \bf IP \\  \bf PSNR\end{tabular} & \begin{tabular}{@{}c@{}} \bf IP \\  \bf SSIM\end{tabular} & \begin{tabular}{@{}c@{}} \bf SOS \\ \bf PSNR\end{tabular} & \begin{tabular}{@{}c@{}} \bf SOS \\ \bf SSIM\end{tabular} & \bf Time \\
            \midrule
            Conventional DAS & 21.49 & 0.372 & - & - & 0.13 s \\
            Dual-SOS DAS~\cite{li2017single} & 24.42 & 0.446 & - & - & 0.12 s \\ 
            APACT~\cite{cui2021adaptive} & 21.49 & 0.434 & 17.74 & 0.908 & 23 min \\
            Ours (PG) & 25.05 & 0.514 & 21.26 & 0.903 & 113.3 s  \\
            Ours (NF) & 25.08 & 0.519 & 22.29 & 0.931 & 40.3 s \\
            Ours (SOS oracle) & 25.61 & 0.537 & - & - & 2.6 s \\
            \bottomrule
        \end{tabular}
        \caption{{\bf Evaluation of reconstruction performance.} Averaged over 5 numerical phantoms. See~\cref{fig:numerical_supp} for illustrations.}
        \label{tab:results}
    \label{tab:ablation}
\end{table}

\subsection{Leaf Phantom}

We use the phantom data from~\cite{cui2021adaptive}, which features a leaf in a 2.5 cm diameter agarose cylinder with a water-filled hole in the middle for acoustic heterogeneity (see photograph in lower left of~\cref{fig:real}a and details in~\cref{sec:leaf_supp}). 
We show the EIR and MTF of our ring array system in and~\cref{sec:eir_mtf}.
The reconstructions of our methods and the baselines are shown in~\cref{fig:real}a. The lower left of the phantom is dim due to nonuniform illumination~\cite{cui2021adaptive}. We searched over SOS values to report the best outputs, using SOS values of 1505.0 m/s for conventional DAS and 1520.0 m/s for body SOS in Dual-SOS DAS. Ours are able to refocus the leaf veins in the cylinder (blue) as well as inside the hole (red), where the signals are weak and DAS and APACT both fail. The SOS reconstructed by our methods have a lower SOS region in the middle matching the hole filled with water, while APACT smooths out the SOS and cannot resolve the hole. PG still suffers from stripe-like artifacts in the SOS (zoom in) but also reconstructs a well-resolved IP image. 

\subsection{In vivo Mouse Liver}

In our {\em in vivo} experiment, we used a nude mouse liver from~\cite{cai2019feature} (details in~\cref{sec:in_vivo_supp}). The reconstructions are shown in~\cref{fig:real}b along with a cartoon illustration of the imaged section (bottom left corner). We hand-tuned SOS assumptions to 1517.0 m/s for conventional DAS and 1560.0 m/s for Dual-SOS DAS.
Despite lacking ground truth, we observe success on several important features. Our method with NF is able to increase the overall contrast and recover structures like blood vessels (green and blue) and bright spine and rib structures (red) in the IP image. Our reconstructed SOS has a superior match to the liver's anatomical shape (see overlaid figure in~\cref{fig:overlay}). In comparison, APACT reconstructs a smooth SOS in which the middle, high-SOS region does not align well with the liver. Ours with PG also suffers from severe artifacts in the SOS.

\subsection{Ablation Studies}
\label{sec:ablations}

\begin{figure}
    \centerline{\includegraphics[width=\columnwidth]{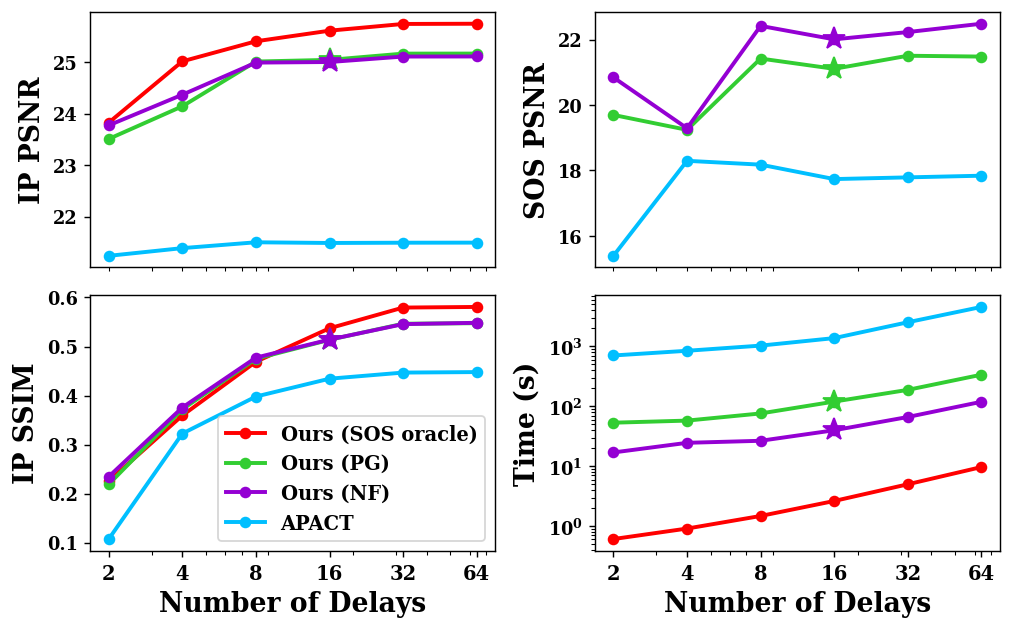}}
    \caption{\textbf{Ablation study on number of delays.} Averaged over 5 numerical phantoms, our method outperforms APACT in both reconstruction quality and compute time. We have shown results for 16 delays but demonstrate here a well-behaved trade-off space.
    }
    \label{fig:ablation_delays}
\end{figure}

Our ablation studies focus on the effects of the number of delays $M$, TV regularization $\lambda$, and network size. 
We observe an increase in the performance at the cost of compute time as more extra delay distances are used and more redundancy is created in the DAS stack. This improvement gradually diminishes beyond a delay count of 16, which we selected for our implementation (see \cref{fig:ablation_delays}). 
For NF network size (see~\cref{fig:ablation_network} and~\cref{sec:ablation_network} in the supplement), our method fails either when the network is too small (NF's implicit smoothness regularization is too strong) or large (the NF has a stronger ability to learn high-frequency features and has weaker implicit regularization). 
For the TV regularization of PG (see~\cref{fig:ablation_tv} and~\cref{sec:ablation_tv} in the supplement), a larger $\lambda$ encourages smoother SOS results and reaches optimal performance at a weight of $1\times10^{-4}$.
\section{Discussion and Conclusion}
\label{sec:discussion}
We have introduced a self-supervised framework achieving fast, state-of-the-art performance in joint reconstruction of initial pressure and speed of sound in PACT. This method combines the neural field representation from computer graphics with an extension of the optical metaphors of autofocus and adaptive optics used in previous methods to a multi-channel image representation analogous to a focal stack. We highlight the following as the key features of our novel method:

\textbf{Reliability and interpretability.} Our method combines signal processing with neural networks to provide reasonable guess of the SOS without a black box. In contrast to deep learning methods where neural networks are trained on large datasets, our method utilizes a small MLP as an efficient representation in a physics-based framework and incorporates redundant information in the measurements to efficiently solve a partially-blind inverse problem. Our approach enhances reliability and interpretability, making it well-suited for scientific and medical applications where black-box models are often undesirable.

\textbf{Time-performance trade-offs.} Existing reconstruction methods offer diverse solutions for different demands. Based on our benchmarking, we can recommend Dual-SOS DAS with a properly-selected SOS for real-time removal of many aberrations. When image quality is key, we offer a new method to access state-of-the-art reconstruction of the IP and SOS in roughly 40 seconds. The delay count, network size, and number of training epochs in NF further allow for customization of the time-performance trade-off. Additionally, we demonstrate 3-second reconstruction when SOS is measured directly.


\textbf{Scalability} Our efficient combination of coordinate-based SOS representations with a differentiable forward model offers a promising solution for large photoacoustic datasets. By reducing the degrees of freedom and lowering computational costs, the proposed method could enable practical large-scale PACT applications.

{
    \section*{Acknowledgments}

We gratefully acknowledge the support of the NSF-Simons AI-Institute for the Sky (SkAI) via grants NSF AST-2421845 and Simons Foundation MPS-AI-00010513.
We would also like to thank Liujie Gu and Yan Luo for helping us with the numerical simulations and Yi-Chun Hung and Marcos Ferreira for suggestions on the manuscript.

    \small\bibliographystyle{ieeenat_fullname}
    \bibliography{main}

\begin{thebibliography}{47}
\providecommand{\natexlab}[1]{#1}
\providecommand{\url}[1]{\texttt{#1}}
\expandafter\ifx\csname urlstyle\endcsname\relax
  \providecommand{\doi}[1]{doi: #1}\else
  \providecommand{\doi}{doi: \begingroup \urlstyle{rm}\Url}\fi

\bibitem[Anastasio et~al.(2005)Anastasio, Zhang, Pan, Zou, Ku, and Wang]{anastasio2005half}
Mark~A Anastasio, Jin Zhang, Xiaochuan Pan, Yu Zou, Geng Ku, and Lihong~V Wang.
\newblock Half-time image reconstruction in thermoacoustic tomography.
\newblock \emph{IEEE transactions on medical imaging}, 24\penalty0 (2):\penalty0 199--210, 2005.

\bibitem[Byra et~al.(2024)Byra, Jarosik, Karwat, Klimonda, and Lewandowski]{byra2024implicit}
Michal Byra, Piotr Jarosik, Piotr Karwat, Ziemowit Klimonda, and Marcin Lewandowski.
\newblock Implicit neural representations for speed-of-sound estimation in ultrasound.
\newblock In \emph{2024 IEEE Ultrasonics, Ferroelectrics, and Frequency Control Joint Symposium (UFFC-JS)}, pages 1--4. IEEE, 2024.

\bibitem[Cai et~al.(2019)Cai, Wang, Si, Qian, Luo, and Ma]{cai2019feature}
Chuangjian Cai, Xuanhao Wang, Ke Si, Jun Qian, Jianwen Luo, and Cheng Ma.
\newblock Feature coupling photoacoustic computed tomography for joint reconstruction of initial pressure and sound speed in vivo.
\newblock \emph{Biomedical optics express}, 10\penalty0 (7):\penalty0 3447--3462, 2019.

\bibitem[Cong et~al.(2015)Cong, Kondo, Namita, Yamakawa, and Shiina]{cong2015photoacoustic}
Bing Cong, Kengo Kondo, Takeshi Namita, Makoto Yamakawa, and Tsuyoshi Shiina.
\newblock Photoacoustic image quality enhancement by estimating mean sound speed based on optimum focusing.
\newblock \emph{Japanese Journal of Applied Physics}, 54\penalty0 (7S1):\penalty0 07HC13, 2015.

\bibitem[Cui et~al.(2021)Cui, Zuo, Wang, Deng, Luo, and Ma]{cui2021adaptive}
Manxiu Cui, Hongzhi Zuo, Xuanhao Wang, Kexin Deng, Jianwen Luo, and Cheng Ma.
\newblock Adaptive photoacoustic computed tomography.
\newblock \emph{Photoacoustics}, 21:\penalty0 100223, 2021.

\bibitem[Deng et~al.(2021)Deng, Qiao, Dai, and Ma]{deng2021deep}
Handi Deng, Hui Qiao, Qionghai Dai, and Cheng Ma.
\newblock Deep learning in photoacoustic imaging: a review.
\newblock \emph{Journal of Biomedical Optics}, 26\penalty0 (4):\penalty0 040901--040901, 2021.

\bibitem[Feng et~al.(2023)Feng, Guo, Xie, Boominathan, Sharma, Veeraraghavan, and Metzler]{feng2023neuws}
Brandon~Y Feng, Haiyun Guo, Mingyang Xie, Vivek Boominathan, Manoj~K Sharma, Ashok Veeraraghavan, and Christopher~A Metzler.
\newblock Neuws: Neural wavefront shaping for guidestar-free imaging through static and dynamic scattering media.
\newblock \emph{Science Advances}, 9\penalty0 (26):\penalty0 eadg4671, 2023.

\bibitem[Fridovich-Keil et~al.(2022)Fridovich-Keil, Yu, Tancik, Chen, Recht, and Kanazawa]{fridovich2022plenoxels}
Sara Fridovich-Keil, Alex Yu, Matthew Tancik, Qinhong Chen, Benjamin Recht, and Angjoo Kanazawa.
\newblock Plenoxels: Radiance fields without neural networks.
\newblock In \emph{Proceedings of the IEEE/CVF conference on computer vision and pattern recognition}, pages 5501--5510, 2022.

\bibitem[Hampson et~al.(2021)Hampson, Turcotte, Miller, Kurokawa, Males, Ji, and Booth]{hampson2021adaptive}
Karen~M Hampson, Rapha{\"e}l Turcotte, Donald~T Miller, Kazuhiro Kurokawa, Jared~R Males, Na Ji, and Martin~J Booth.
\newblock Adaptive optics for high-resolution imaging.
\newblock \emph{Nature Reviews Methods Primers}, 1\penalty0 (1):\penalty0 68, 2021.

\bibitem[Hoelen and de~Mul(2000)]{hoelen2000image}
Christoph~GA Hoelen and Frits~FM de Mul.
\newblock Image reconstruction for photoacoustic scanning of tissue structures.
\newblock \emph{Applied Optics}, 39\penalty0 (31):\penalty0 5872--5883, 2000.

\bibitem[Jeon and Kim(2020)]{jeon2020deep}
Seungwan Jeon and Chulhong Kim.
\newblock Deep learning-based speed of sound aberration correction in photoacoustic images.
\newblock In \emph{Photons plus ultrasound: Imaging and sensing 2020}, pages 24--27. SPIE, 2020.

\bibitem[Jeon et~al.(2019)Jeon, Park, Choi, Managuli, jong Lee, and Kim]{jeon2019real}
Seungwan Jeon, Eun-Yeong Park, Wonseok Choi, Ravi Managuli, Ki jong Lee, and Chulhong Kim.
\newblock Real-time delay-multiply-and-sum beamforming with coherence factor for in vivo clinical photoacoustic imaging of humans.
\newblock \emph{Photoacoustics}, 15:\penalty0 100136, 2019.

\bibitem[Jeon et~al.(2021)Jeon, Choi, Park, and Kim]{jeon2021deep}
Seungwan Jeon, Wonseok Choi, Byullee Park, and Chulhong Kim.
\newblock A deep learning-based model that reduces speed of sound aberrations for improved in vivo photoacoustic imaging.
\newblock \emph{IEEE transactions on image processing}, 30:\penalty0 8773--8784, 2021.

\bibitem[Ji(2017)]{ji2017adaptive}
Na Ji.
\newblock Adaptive optical fluorescence microscopy.
\newblock \emph{Nature methods}, 14\penalty0 (4):\penalty0 374--380, 2017.

\bibitem[Jin and Wang(2006)]{jin2006thermoacoustic}
Xing Jin and Lihong~V Wang.
\newblock Thermoacoustic tomography with correction for acoustic speed variations.
\newblock \emph{Physics in Medicine \& Biology}, 51\penalty0 (24):\penalty0 6437, 2006.

\bibitem[Kang et~al.(2024)Kang, Zhang, Yu, and Ji]{kang2024coordinate}
Iksung Kang, Qinrong Zhang, Stella~X Yu, and Na Ji.
\newblock Coordinate-based neural representations for computational adaptive optics in widefield microscopy.
\newblock \emph{Nature Machine Intelligence}, 6\penalty0 (6):\penalty0 714--725, 2024.

\bibitem[Kim and Fridovich-Keil(2025)]{kim2025grids}
Namhoon Kim and Sara Fridovich-Keil.
\newblock Grids often outperform implicit neural representations.
\newblock \emph{arXiv preprint arXiv:2506.11139}, 2025.

\bibitem[Kingma(2014)]{kingma2014adam}
Diederik~P Kingma.
\newblock Adam: A method for stochastic optimization.
\newblock \emph{arXiv preprint arXiv:1412.6980}, 2014.

\bibitem[Li et~al.(2017)Li, Zhu, Ma, Lin, Yao, Wang, Maslov, Zhang, Chen, Shi, et~al.]{li2017single}
Lei Li, Liren Zhu, Cheng Ma, Li Lin, Junjie Yao, Lidai Wang, Konstantin Maslov, Ruiying Zhang, Wanyi Chen, Junhui Shi, et~al.
\newblock Single-impulse panoramic photoacoustic computed tomography of small-animal whole-body dynamics at high spatiotemporal resolution.
\newblock \emph{Nature biomedical engineering}, 1\penalty0 (5):\penalty0 0071, 2017.

\bibitem[Manohar et~al.(2007)Manohar, Willemink, van~der Heijden, Slump, and van Leeuwen]{manohar2007concomitant}
Srirang Manohar, Ren{\'e}~GH Willemink, Ferdi van~der Heijden, Cornelis~H Slump, and Ton~G van Leeuwen.
\newblock Concomitant speed-of-sound tomography in photoacoustic imaging.
\newblock \emph{Applied physics letters}, 91\penalty0 (13), 2007.

\bibitem[Marczak(1997)]{marczak1997water}
Wojciech Marczak.
\newblock Water as a standard in the measurements of speed of sound in liquids.
\newblock \emph{the Journal of the Acoustical Society of America}, 102\penalty0 (5):\penalty0 2776--2779, 1997.

\bibitem[Molaei et~al.(2023)Molaei, Aminimehr, Tavakoli, Kazerouni, Azad, Azad, and Merhof]{molaei2023implicit}
Amirali Molaei, Amirhossein Aminimehr, Armin Tavakoli, Amirhossein Kazerouni, Bobby Azad, Reza Azad, and Dorit Merhof.
\newblock Implicit neural representation in medical imaging: A comparative survey.
\newblock In \emph{Proceedings of the IEEE/CVF International Conference on Computer Vision}, pages 2381--2391, 2023.

\bibitem[Paszke et~al.(2019)Paszke, Gross, Massa, Lerer, Bradbury, Chanan, Killeen, Lin, Gimelshein, Antiga, et~al.]{paszke2019pytorch}
Adam Paszke, Sam Gross, Francisco Massa, Adam Lerer, James Bradbury, Gregory Chanan, Trevor Killeen, Zeming Lin, Natalia Gimelshein, Luca Antiga, et~al.
\newblock Pytorch: An imperative style, high-performance deep learning library.
\newblock \emph{Advances in neural information processing systems}, 32, 2019.

\bibitem[Poudel et~al.(2017)Poudel, Matthews, Li, Anastasio, and Wang]{poudel2017mitigation}
Joemini Poudel, Thomas~P Matthews, Lei Li, Mark~A Anastasio, and Lihong~V Wang.
\newblock Mitigation of artifacts due to isolated acoustic heterogeneities in photoacoustic computed tomography using a variable data truncation-based reconstruction method.
\newblock \emph{Journal of biomedical optics}, 22\penalty0 (4):\penalty0 041018--041018, 2017.

\bibitem[Reed et~al.(2021)Reed, Kim, Anirudh, Mohan, Champley, Kang, and Jayasuriya]{reed2021dynamic}
Albert~W Reed, Hyojin Kim, Rushil Anirudh, K~Aditya Mohan, Kyle Champley, Jingu Kang, and Suren Jayasuriya.
\newblock Dynamic ct reconstruction from limited views with implicit neural representations and parametric motion fields.
\newblock In \emph{Proceedings of the IEEE/CVF International Conference on Computer Vision}, pages 2258--2268, 2021.

\bibitem[Shan et~al.(2019)Shan, Wiedeman, Wang, and Yang]{shan2019simultaneous}
Hongming Shan, Christopher Wiedeman, Ge Wang, and Yang Yang.
\newblock Simultaneous reconstruction of the initial pressure and sound speed in photoacoustic tomography using a deep-learning approach.
\newblock In \emph{Novel Optical Systems, Methods, and Applications XXII}, pages 18--27. SPIE, 2019.

\bibitem[Shen et~al.(2022)Shen, Pauly, and Xing]{shen2022nerp}
Liyue Shen, John Pauly, and Lei Xing.
\newblock Nerp: implicit neural representation learning with prior embedding for sparsely sampled image reconstruction.
\newblock \emph{IEEE Transactions on Neural Networks and Learning Systems}, 35\penalty0 (1):\penalty0 770--782, 2022.

\bibitem[Sitzmann et~al.(2020)Sitzmann, Martel, Bergman, Lindell, and Wetzstein]{sitzmann2020implicit}
Vincent Sitzmann, Julien Martel, Alexander Bergman, David Lindell, and Gordon Wetzstein.
\newblock Implicit neural representations with periodic activation functions.
\newblock \emph{Advances in neural information processing systems}, 33:\penalty0 7462--7473, 2020.

\bibitem[Sun et~al.(2021)Sun, Liu, Xie, Wohlberg, and Kamilov]{sun2021coil}
Yu Sun, Jiaming Liu, Mingyang Xie, Brendt Wohlberg, and Ulugbek~S Kamilov.
\newblock Coil: Coordinate-based internal learning for tomographic imaging.
\newblock \emph{IEEE Transactions on Computational Imaging}, 7:\penalty0 1400--1412, 2021.

\bibitem[Tang et~al.(2023)Tang, Zhang, Liang, Wang, Ge, Chen, and Qi]{tang2023advanced}
Kaiyi Tang, Shuangyang Zhang, Zhichao Liang, Yang Wang, Jia Ge, Wufan Chen, and Li Qi.
\newblock Advanced image post-processing methods for photoacoustic tomography: A review.
\newblock In \emph{Photonics}, page 707. MDPI, 2023.

\bibitem[Thomenius(1996)]{thomenius1996evolution}
Kai~E Thomenius.
\newblock Evolution of ultrasound beamformers.
\newblock In \emph{1996 IEEE Ultrasonics Symposium. Proceedings}, pages 1615--1622. IEEE, 1996.

\bibitem[Treeby and Cox(2010)]{treeby2010k}
Bradley~E Treeby and Benjamin~T Cox.
\newblock k-wave: Matlab toolbox for the simulation and reconstruction of photoacoustic wave fields.
\newblock \emph{Journal of biomedical optics}, 15\penalty0 (2):\penalty0 021314--021314, 2010.

\bibitem[Treeby et~al.(2011)Treeby, Varslot, Zhang, Laufer, and Beard]{treeby2011automatic}
Bradley~E Treeby, Trond~K Varslot, Edward~Z Zhang, Jan~G Laufer, and Paul~C Beard.
\newblock Automatic sound speed selection in photoacoustic image reconstruction using an autofocus approach.
\newblock \emph{Journal of biomedical optics}, 16\penalty0 (9):\penalty0 090501--090501, 2011.

\bibitem[Wang et~al.(2020)Wang, Liu, and Tian]{wang2020combating}
Tong Wang, Wen Liu, and Chao Tian.
\newblock Combating acoustic heterogeneity in photoacoustic computed tomography: A review.
\newblock \emph{Journal of Innovative Optical Health Sciences}, 13\penalty0 (03):\penalty0 2030007, 2020.

\bibitem[Xia et~al.(2013)Xia, Huang, Maslov, Anastasio, and Wang]{xia2013enhancement}
Jun Xia, Chao Huang, Konstantin Maslov, Mark~A Anastasio, and Lihong~V Wang.
\newblock Enhancement of photoacoustic tomography by ultrasonic computed tomography based on optical excitation of elements of a full-ring transducer array.
\newblock \emph{Optics letters}, 38\penalty0 (16):\penalty0 3140--3143, 2013.

\bibitem[Xia et~al.(2014)Xia, Yao, and Wang]{xia2014photoacoustic}
Jun Xia, Junjie Yao, and Lihong~V Wang.
\newblock Photoacoustic tomography: principles and advances.
\newblock \emph{Electromagnetic waves (Cambridge, Mass.)}, 147:\penalty0 1, 2014.

\bibitem[Xiao et~al.(2024)Xiao, Shen, Yao, Cai, and Gao]{xiao2024unsupervised}
Youshen Xiao, Yuting Shen, Bowei Yao, Xiran Cai, and Fei Gao.
\newblock Unsupervised neural representation for limited-view photoacoustic imaging reconstruction.
\newblock In \emph{2024 IEEE Ultrasonics, Ferroelectrics, and Frequency Control Joint Symposium (UFFC-JS)}, pages 1--3. IEEE, 2024.

\bibitem[Xie et~al.(2022)Xie, Takikawa, Saito, Litany, Yan, Khan, Tombari, Tompkin, Sitzmann, and Sridhar]{xie2022neural}
Yiheng Xie, Towaki Takikawa, Shunsuke Saito, Or Litany, Shiqin Yan, Numair Khan, Federico Tombari, James Tompkin, Vincent Sitzmann, and Srinath Sridhar.
\newblock Neural fields in visual computing and beyond.
\newblock In \emph{Computer Graphics Forum}, pages 641--676. Wiley Online Library, 2022.

\bibitem[Xu et~al.(2023)Xu, Moyer, Gagoski, Iglesias, Grant, Golland, and Adalsteinsson]{xu2023nesvor}
Junshen Xu, Daniel Moyer, Borjan Gagoski, Juan~Eugenio Iglesias, P~Ellen Grant, Polina Golland, and Elfar Adalsteinsson.
\newblock Nesvor: implicit neural representation for slice-to-volume reconstruction in mri.
\newblock \emph{IEEE transactions on medical imaging}, 42\penalty0 (6):\penalty0 1707--1719, 2023.

\bibitem[Xu and Wang(2005)]{xu2005universal}
Minghua Xu and Lihong~V Wang.
\newblock Universal back-projection algorithm for photoacoustic computed tomography.
\newblock \emph{Physical Review E—Statistical, Nonlinear, and Soft Matter Physics}, 71\penalty0 (1):\penalty0 016706, 2005.

\bibitem[Xu and Wang(2003)]{xu2003effects}
Yuan Xu and Lihong~V Wang.
\newblock Effects of acoustic heterogeneity in breast thermoacoustic tomography.
\newblock \emph{IEEE transactions on ultrasonics, ferroelectrics, and frequency control}, 50\penalty0 (9):\penalty0 1134--1146, 2003.

\bibitem[Yao et~al.(2024)Yao, Cui, Dai, Wu, Xiao, Gao, Yu, Zhang, and Cai]{yao2024implicit}
Bowei Yao, Shilong Cui, Haizhao Dai, Qing Wu, Youshen Xiao, Fei Gao, Jingyi Yu, Yuyao Zhang, and Xiran Cai.
\newblock Implicit neural representation for sparse-view photoacoustic computed tomography.
\newblock \emph{arXiv preprint arXiv:2409.13696}, 2024.

\bibitem[Yoon et~al.(2012)Yoon, Kang, Han, Yoo, Song, and Chang]{yoon2012enhancement}
Changhan Yoon, Jeeun Kang, Seunghee Han, Yangmo Yoo, Tai-Kyong Song, and Jin~Ho Chang.
\newblock Enhancement of photoacoustic image quality by sound speed correction: ex vivo evaluation.
\newblock \emph{Optics express}, 20\penalty0 (3):\penalty0 3082--3090, 2012.

\bibitem[Zang et~al.(2021)Zang, Idoughi, Li, Wonka, and Heidrich]{zang2021intratomo}
Guangming Zang, Ramzi Idoughi, Rui Li, Peter Wonka, and Wolfgang Heidrich.
\newblock Intratomo: self-supervised learning-based tomography via sinogram synthesis and prediction.
\newblock In \emph{Proceedings of the IEEE/CVF International Conference on Computer Vision}, pages 1960--1970, 2021.

\bibitem[Zhang and Anastasio(2006)]{zhang2006reconstruction}
Jin Zhang and Mark~A Anastasio.
\newblock Reconstruction of speed-of-sound and electromagnetic absorption distributions in photoacoustic tomography.
\newblock In \emph{Photons Plus Ultrasound: Imaging and Sensing 2006: The Seventh Conference on Biomedical Thermoacoustics, Optoacoustics, and Acousto-optics}, pages 339--345. SPIE, 2006.

\bibitem[Zhang et~al.(2023)Zhang, Hu, Berlage, Kner, Judkewitz, Booth, and Ji]{zhang2023adaptive}
Qinrong Zhang, Qi Hu, Caroline Berlage, Peter Kner, Benjamin Judkewitz, Martin Booth, and Na Ji.
\newblock Adaptive optics for optical microscopy.
\newblock \emph{Biomedical Optics Express}, 14\penalty0 (4):\penalty0 1732--1756, 2023.

\bibitem[Zou et~al.(2024)Zou, Lin, and Zhu]{zou2024pa}
Yun Zou, Yixiao Lin, and Quing Zhu.
\newblock Pa-nerf, a neural radiance field model for 3d photoacoustic tomography reconstruction from limited bscan data.
\newblock \emph{Biomedical Optics Express}, 15\penalty0 (3):\penalty0 1651--1667, 2024.

\end{thebibliography}
}

\newcounter{si}
\setcounter{si}{0} 
\renewcommand\thesection{S\arabic{si}}
\newcounter{fi}
\setcounter{fi}{0} 
\renewcommand{\thefigure}{S\arabic{fi}}

\clearpage
\setcounter{page}{1}
\maketitlesupplementary


\stepcounter{si}
\section{Experiment Details}
\label{sec:exp_details}

\subsection{Numerical Phantoms}
\label{sec:simulations_supp}

In this section, we describe the simulation details of our numerical phantoms.

\begin{figure*}
    \centerline{\includegraphics[width=1.9\columnwidth]{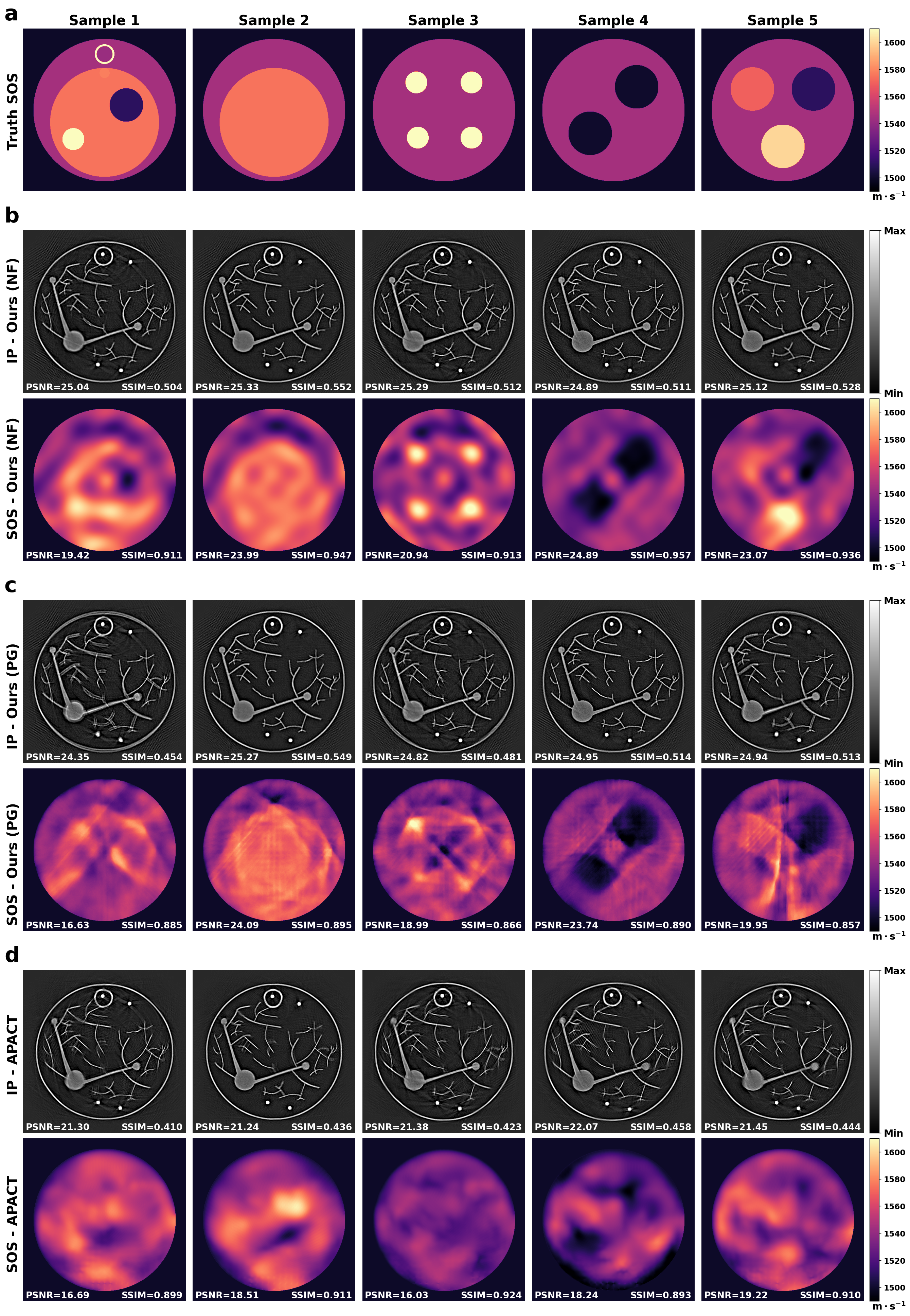}}
    \caption{\textbf{All five numerical phantoms and results.} 
    (a) The SOS maps used in our numerical phantoms.
    (b) Reconstructions of our method with NF.
    (c) Reconstructions of our method with PG.
    (d) Reconstructions of APACT~\cite{cui2021adaptive}.
    }
    \label{fig:numerical_supp}
\end{figure*}

We use the same IP image shown in~\cref{fig:numerical}a for all numerical phantoms but use five SOS with different variations (see~\cref{fig:numerical_supp}a). SOS values range from 1490.0 to 1650.0 m/s in our numerical dataset. The structures in the IP and SOS images are inspired by the mouse body\footnote{\href{https://www.imaios.com/en/vet-anatomy/mouse/mouse-whole-body}{https://www.imaios.com/en/vet-anatomy/mouse/mouse-whole-body}} while the SOS values are based on realistic measurements\footnote{\href{https://itis.swiss/virtual-population/tissue-properties/database/acoustic-properties/}{https://itis.swiss/virtual-population/tissue-properties/database/acoustic-properties/}}. For example, the small circles with high SOS in samples 1 and 3 correspond to bones, the orange circles in samples 1 and 2 correspond to the liver, and the circle with small SOS in samples 1, 4, and 5 correspond to the stomach lumen.

The background water SOS is set to 1499.4 m/s (assuming a 26$^{\circ}$C water temperature based on~\cite{marczak1997water}) outside the sample. The PA signals $p(n,t)$ are obtained by 2D numerical simulation using the k-Wave toolbox~\cite{treeby2010k} assuming an evenly distributed 512-transducer ring array with a diameter of 10 cm (same setting for the leaf phantom and {\em in vivo} mouse liver experiments). Similar to~\cite{cui2021adaptive}, we take the derivative of the PA signals to simulate transducer effects: $S(n,t)=-2\frac{\partial p(n,t)}{\partial t}$. The images are cropped into 75\% overlapping 3.2mm $\times$ 3.2mm patches, as for all reported results.

We show the IP and SOS reconstructions on all five numerical phantoms using our methods and APACT in~\cref{fig:numerical_supp} b,c,d. Results are qualitatively similar to the example discussed in the text, and numbers reported in tables are averaged over these five phantoms.

\subsection{Leaf Phantom}
\label{sec:leaf_supp}

We use the experimentally collected phantom data from~\cite{cui2021adaptive}, which features a leaf in a 2.5 cm diameter agarose cylinder with a water-filled hole in the middle for acoustic heterogeneity (see photograph in \cref{fig:real}a). 
The leaf phantom is surrounded by water with a temperature of 26$^{\circ}$C, and the wavelength of laser illumination is 700 nm in the experiment. 
The constant background SOS is set to 1499.4 m/s according to~\cite{marczak1997water}. 

Due to nonideal transducer alignment and transducer impulse responses, the collected PA signals are calibrated using measured ring error and transducer electrical impulse response (EIR) before feeding into the algorithms. The same calibration is applied to the {\em in vivo} experiment.

\subsection{In vivo Mouse Liver}
\label{sec:in_vivo_supp}
\begin{figure*}[t]
    \centerline{\includegraphics[width=2\columnwidth]{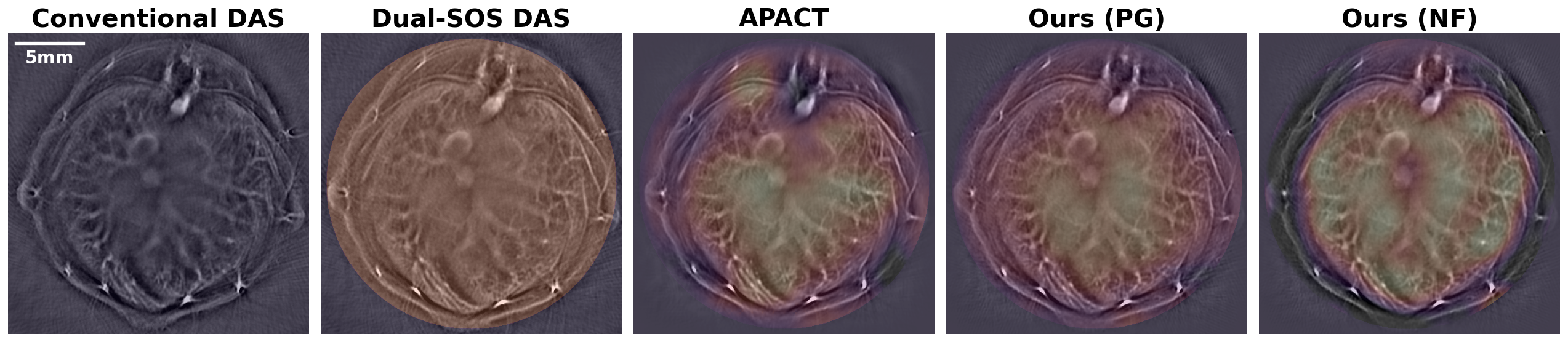}}
    \caption{\textbf{Overlaid initial pressure image and SOS of {\em in vivo} mouse liver.} 
        Our reconstructed SOS with NF (right) shows a superior match to the liver’s anatomical shape in the initial pressure image.
    }
    \label{fig:overlay}
\end{figure*}

The PA signals in our {\em in vivo} experiments were collected from a nude mouse liver at 1064 nm~\cite{cai2019feature}. 
The water temperature was 31$^{\circ}$C during the experiment, thus we used a background SOS of 1511.4 m/s. See~\cref{fig:overlay} to see the alignment between recovered SOS and IP, which matches better in our method than others.

\stepcounter{fi}
\begin{figure}[b]
    \centerline{\includegraphics[width=\columnwidth]{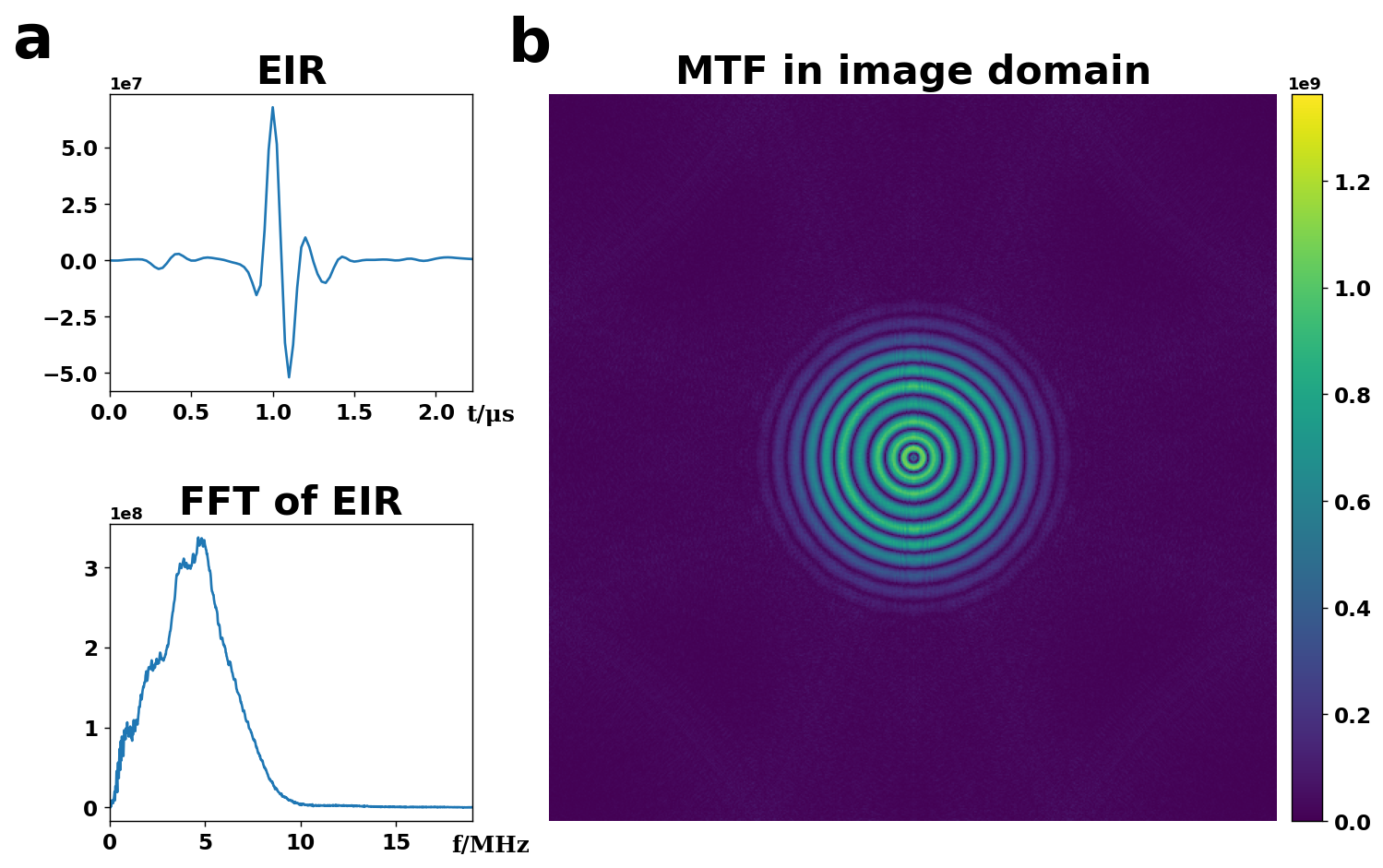}}
    \caption{
    (a) Experimentally measured transducer EIR and its Fourier transform.
    (b) The system MTF is the Fourier transform of the PSF, which is the projection of EIR into image space.
    }
    \label{fig:mtf}
\end{figure}

\subsection{Transducer EIR and System MTF}
\label{sec:eir_mtf}

We measured the transducer electrical impulse response (EIR) in the setup we used to collect the data of the leaf phantom and the {\em in vivo} mouse liver.
We show that our experimentally collected transducer EIR and its Fourier transform is a band-pass filter, shown in~\cref{fig:mtf}a .
We calculate the system PSF by projecting this EIR from all 512 transducers into the image space using DAS and calculate the system modulation transfer function (MTF) by taking the 2D Fourier transform of this PSF, shown in ~\cref{fig:mtf}b.
The ring structure in the MTF is caused by the projection arcs of DAS.
We apply a phase only EIR deconvolution on the collected PA signals before feeding them into our algorithm as a calibration in the phantom and the {\em in vivo} experiment.

\stepcounter{fi}

\subsection{Image Stitching}
\label{sec:image_stichting}

We adopted the same strategy used for APACT~\cite{cui2021adaptive} in IP reconstruction. This process differs from most existing image reconstruction frameworks because the image is reconstructed patch-by-patch and stitched to form the final result. Images are cropped using 3.2 mm $\times$ 3.2 mm Gaussian windows with an FWHM of 1.5 mm before being fed to our algorithm. The patch centers are arranged on a Cartesian grid, and adjacent patches overlap by 75\%. With our proposed method for solving the speed of sound distribution by utilizing information from all patches, we obtain the wavefront correction model for each isoplanatic patch. This simultaneously yields the deconvolved, aberration-corrected image patches. These patches are then translated to their corresponding center positions and summed. The 75\% overlap ensures a smooth transition during image stitching. Finally, the result is adjusted to remove the modulation pattern due to the Gaussian windows by dividing by the sum of the Gaussian window weights.

\stepcounter{si}
\section{Difference Between Our Wavefront Model and APACT's Wavefront Model}
\label{sec:apact_wavefront}

One critical contribution of our work is the improved wavefront model compared to the previous work APACT~\cite{cui2021adaptive}. As explained in the main text, the wavefront model represents the wavefront advancement or retardation induced by a nonuniform speed of sound in tissue relative to an ideal spherical wave in a homogeneous medium. Mathematically, the wavefront function \(w(\theta)\), which depicts the propagation length difference relative to the spherical wave when received by the transducers, is a function of the propagation direction \(\theta\). This wavefront function can be expanded using a Fourier series:
$$w(\theta) = C + \sum^\infty_{n=1}A_n\cos(n\theta) + B_n\sin(n\theta).$$
In APACT, only three low order harmonics were used, namely DC (constant $C$), \(\cos(2\theta)\), and \(\sin(2\theta)\). This is because, firstly, the first order terms \(\cos(\theta)\) and \(\sin(\theta)\) result in a shifted PSF, making them non-identifiable from the reconstructed images when the true IP is unknown. Secondly, the authors omitted the higher order terms, i.e. \(\cos(n\theta)\) and \(\sin(n\theta)\) for \(n \geq 3\), to avoid the exponential increase in computation time during the exhaustive search for the optimal wavefront. However, as shown in Fig.~\ref{fig:SOS}, the actual wavefront function can be complex, which necessitate those high order terms. In our work, the wavefront functions corresponding to different isoplanatic patches are obtained from the learned SOS map. There is no explicit restriction on their profiles such as the assumption in APACT that they are slowly varying. 

\stepcounter{si}
\section{Ablation Studies}
\label{sec:ablations_supp}

We conducted ablation studies using our synthetic numerical phantoms on an NVIDIA RTX A6000 GPU.
All results are averaged over 5 numerical phantoms, which share the same IP but have different SOS.

\stepcounter{fi}
\begin{figure}[!b]
    \centerline{\includegraphics[width=\columnwidth]{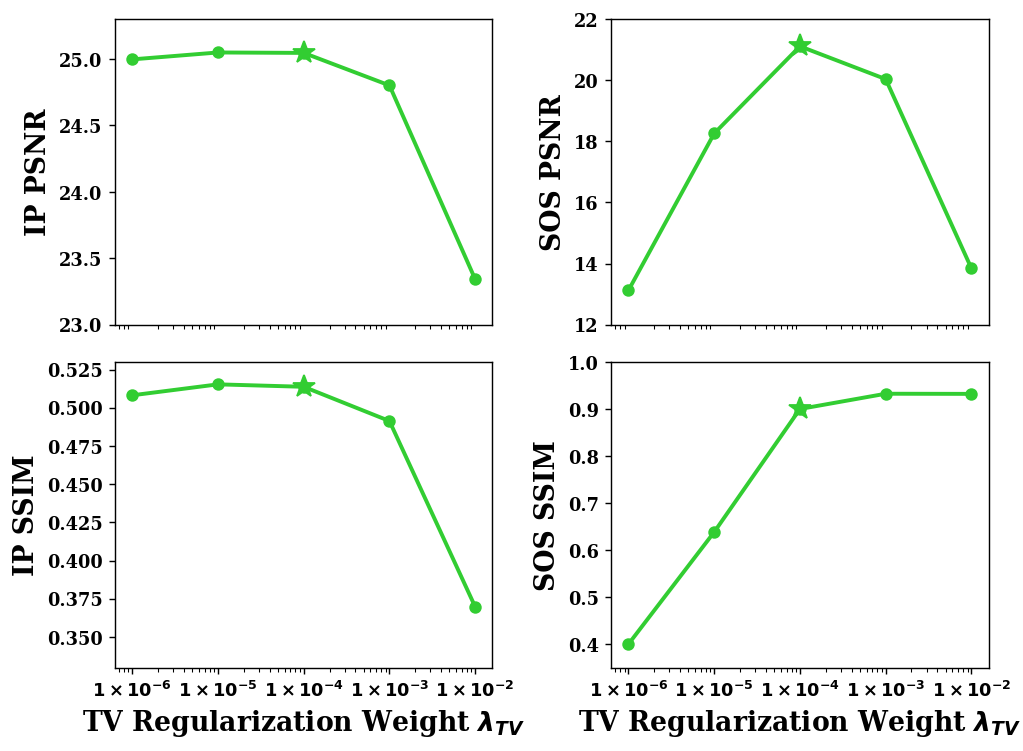}}
    \caption{\textbf{Ablation results on TV regularization weight $\lambda$ in \cref{eq:jr}.} 
    We select $\lambda=1\times10^{-4}$ for the best IP and SOS reconstruction (marked with a green star).
    }
    \label{fig:ablation_tv}
\end{figure}

\subsection{Number of Delays}

\stepcounter{fi}
\begin{figure*}[h!]
    \centerline{\includegraphics[width=1.62\columnwidth]{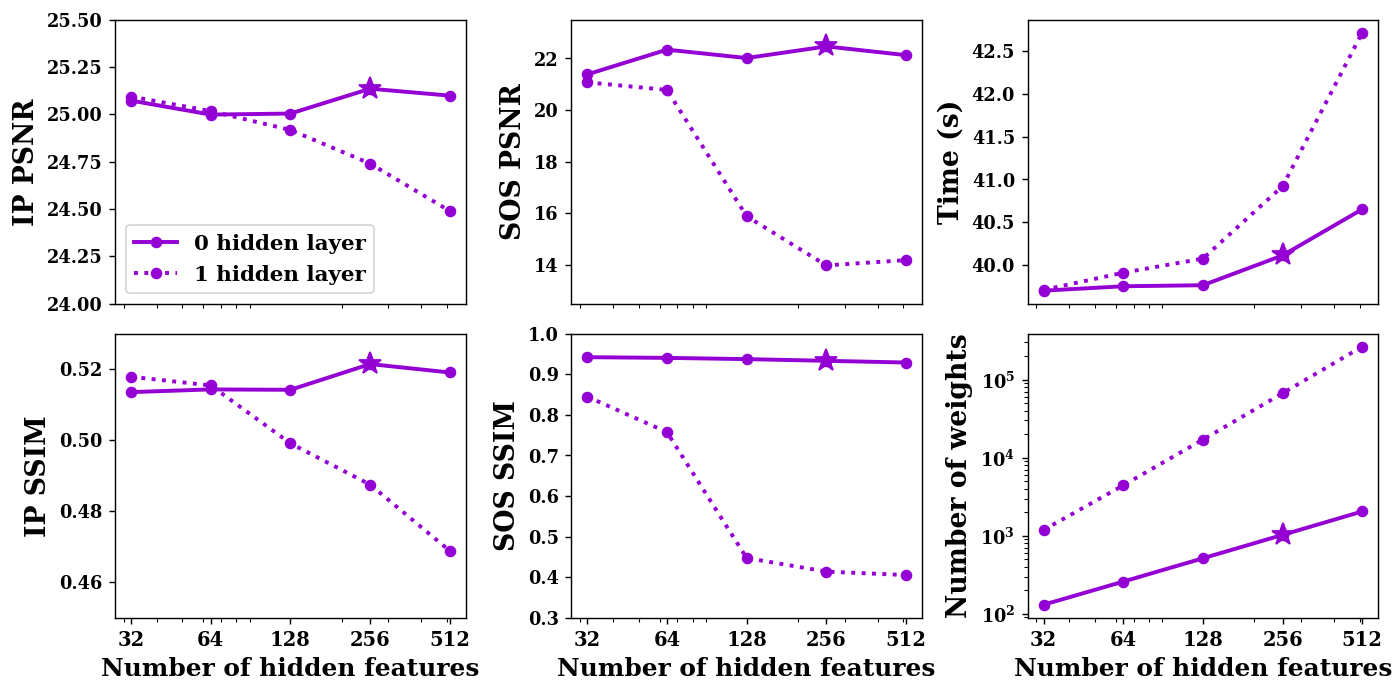}}
    \caption{\textbf{Ablation study on the network size of NF (averaged over 5 numerical phantoms).} 
    Larger network size decreases the implicit smoothness regularization on the SOS, resulting in noisier SOS and IP reconstructions.
    We select 0 hidden layer with 256 hidden features (marked with a purple star).
    }
    \label{fig:ablation_network}
\end{figure*}

We show the performance of our method (with and without SOS oracle), pixel grid (with and without TV regularization), and APACT with respect to numbers of delays used in~\cref{fig:ablation_delays}.
We keep the range of the delay distances fixed at [-0.8 mm, 0.8 mm] and uniformly sample delay distances within this range based on the number of delays used.
We show that more delay channels leads to increased performance and computation time for all methods.
This originates from~\cref{eq:multi-deconv}, where more delay channels promote the performance of pseudo-inverse deconvolution, provide more redundant information for SOS estimation, and require more computation time. 
Removing TV regularization in pixel grid (dashed green line) has little impact on the quality of the IP image but significantly decreases the quality of the SOS. 
Considering the performance and computation time, we choose 16 delays (marked with stars).
We also show that the efficient parameterization of SOS gives NF a slight time advantage compared to PG (see lower right subplot), which takes more iterations to reach convergence.

\subsection{TV regularization for PG}
\label{sec:ablation_tv}

In~\cref{fig:ablation_tv}, we show the effects of TV regularization weight $\lambda$ with our pixel grid SOS representation.
We observe that the IP image quality drops when $\lambda>1\times10^{-4}$, while the SOS accuracy reaches optimum around $\lambda=1\times10^{-4}$. 
Therefore, we select $\lambda=1\times10^{-4}$ (marked with a green star in~\cref{fig:ablation_tv}) for all reconstructions in our implementation of PG.

\subsection{Network Size of NF}
\label{sec:ablation_network}

We compared the performance of our method with NF with different network sizes ranging from 131 to 264,707 parameters, shown in~\cref{fig:ablation_network}. 
We show that the implicit smoothness regularization of the MLP decreases as the network becomes larger, resulting in noisier SOS reconstructions and hence worse IP images.
This can be observed from the comparison between no hidden layer (solid line) and one hidden layer (dashed line) and the overall decreasing trend of the two curves.
An increase in network size also leads to a small rise in computation time.
We select 0 hidden layer and 256 hidden features after an overall consideration of the performance and computation time (marked with a purple star).

\stepcounter{si}
\section{Supplementary Videos}
\label{sec:videos} 

We provide six videos showing the convergence process of the IP image and SOS for our methods with our numerical phantom (\path{nf_numerical_phantom.mp4} and \path{pg_numerical_phantom.mp4}), leaf phantom (\path{nf_leaf_phantom.mp4} and \path{pg_leaf_phantom.mp4}), and {\em in vivo} mouse liver (\path{nf_in_vivo_mouse_liver.mp4} and \path{pg_in_vivo_mouse_liver.mp4}).

\stepcounter{fi}

\end{document}